\documentclass[aps,preprint,showpacs,groupedaddress,floatfix]{revtex4-1}

\usepackage{amssymb}
\usepackage{graphicx}
\usepackage{rotating}
\usepackage{xcolor}
\usepackage{latexsym}
\usepackage{bm}
\usepackage{amsmath}
\begin{document}
\newcommand{\kp}{K^+}
\newcommand{\gk}{\vec{\gamma}\vec{k}}
\newcommand{\gE}{\gamm a_0 E_k}
\newcommand{\ppl}{\vec{p}}
\newcommand{\bcm}{\vec{b}^{\star}}
\newcommand{\becm}{\vec{\beta}^{\star}}
\newcommand{\bepl}{\vec{\beta}}
\newcommand{\rcm}{\vec{r}^{\star}}
\newcommand{\rpl}{\vec{r}}
\newcommand{\A}{{$\mathcal A$}}
\newcommand{\wpk}{ \omega_{p-k}}
\newcommand{\Journal}[4]{ #1 {\bf #2} (#4) #3}
\newcommand{\NPA}{Nucl.\ Phys.\ A}
\newcommand{\PLB}{Phys.\ Lett.\ B}
\newcommand{\PRC}{Phys.\ Rev.\ C}
\newcommand{\ZPC}{Z.\ Phys.\ C}
\newcommand{\be}{\begin{equation}}
\newcommand{\ee}{\end{equation}}

\newcommand{\leftg}{\langle \phi_0 |}
\newcommand{\rightg}{| \phi_0 \rangle}
\newcommand{\bra}[1]{\left\langle #1 \right|}
\newcommand{\ket}[1]{\left| #1 \right\rangle}
\newcommand{\bff}[1]{{\mbox{\boldmath $#1$}}}

\bibliographystyle{prsty}
\title{Relativistic Energy Density Functional Description of Shape Transition in 
Superheavy Nuclei}
\author{V. Prassa$^{1,2}$}
\author{T. Nik\v si\' c $^1$}
\author{G. A. Lalazissis$^2$}
\author{D. Vretenar$^1$}
 \affiliation{$^1$Physics Department, Faculty of science, University of Zagreb, 10000 Zagreb, Croatia}
 \affiliation{$^2$Department of Theoretical Physics,
Aristotle University of Thessaloniki, Thessaloniki Gr-54124, Greece}
\begin{abstract}
Relativistic energy density functionals (REDF) provide a complete and accurate, 
global description of nuclear structure phenomena. A modern semi-empirical 
functional, adjusted to the nuclear matter equation of state and to empirical 
masses of deformed nuclei, is applied to studies of shapes of superheavy 
nuclei. The theoretical framework is tested in a comparison of calculated masses, 
quadrupole deformations, and potential energy barriers to  
available data on actinide isotopes. Self-consistent mean-field calculations 
predict a variety of spherical, axial and triaxial shapes of long-lived superheavy
nuclei, and their alpha-decay energies and half-lives are compared to data. 
A microscopic, REDF-based, quadrupole collective Hamiltonian model is used to 
study the effect of explicit treatment of collective correlations in the calculation of 
$Q_\alpha$ values and half-lives. 
\end{abstract}
\date{\today}
\maketitle

\section{\label{secI} Introduction}
For many years theoretical studies of superheavy nuclei (SHN) were mostly based on the traditional 
macroscopic-microscopic approach 
\cite{Brack.72, MN.94,MNMS.95,Sob.94,SobPom.07}, but since the late 1990s 
the framework of self-consistent mean-field models, based on realistic effective inter-nucleon 
interactions or energy density functionals, has systematically been applied to the structure of SHN  
\cite{SobPom.07,cwiok96,lala.96,Ben.98,Ben.99,cwiok99,Ber.01,War.02,HN.02,Gor.02,BHR.03,Bur.04,CHN.05,DGGL.06,
Sheikh.09,Erler.10,LA.11,Lit.12,Abu.12}. 
Binding energies, deformations, $\alpha$-decay energies 
and half-lives, fission barriers and spontaneous-fission half-lives, fission isomers, and single-nucleon 
shell structure of SHN have successfully been described using self-consistent mean-field (SCMF) 
models based on the Gogny effective interaction, the Skyrme energy functional, and relativistic 
meson-exchange effective Lagrangians.

The advantages of using SCMF models include the 
intuitive interpretation of results in terms of single-particle states 
and intrinsic shapes, calculations performed in 
the full model space of occupied states, and the
universality that enables their applications to all
nuclei throughout the periodic chart. The latter feature is
especially important for extrapolations to regions of exotic
short-lived nuclei far from stability for which few, if any,
data are available. In addition, the SCMF approach can be extended 
beyond the static mean-field level to explicitly include collective 
correlations and, thus, perform detailed calculations of excitation 
spectra and transition rates. 

During the last decade, important experimental results on the 
mass limit of the nuclear chart have been obtained using compound nucleus  
reactions between the $^{48}$Ca beam and actinide targets. A number of isotopes 
of new elements with the atomic number $Z=113-118$ have been dicovered, 
and new isotopes of $Z=110$ and 
112 \cite{og2007,hofmann2007,og2010,og2011,og2012,stavsetra2009,dullmann2010,ellison2010}. 
The decay energies and the resulting half-lives provide evidence of a significant increase 
of stability with increasing neutron number in this region of SHN. Theoretical studies 
predict that SHN in this region should display rapid shape transitions, from prolate, through 
spherical, to oblate-deformed ground states \cite{cwiok96,cwiok99,HN.02,CHN.05,SobPom.07,JKS.11}.
These nuclei, therefore, present an ideal testing ground for structure models attempting to 
predict the location of an ``island of stability" for SHN around $N=184$. 

In this work we apply the framework of relativistic energy density functionals (REDF) to an 
illustrative study of shape transitions and shape coexistence in SHN with $Z=110-120$. 
There are several advantages in using functionals with manifest
covariance and, in the context of this study, the most important is the natural inclusion of the
nucleon spin degree of freedom, and the resulting nuclear spin-orbit potential
which emerges automatically with the empirical strength. Our aim is to test the recently 
introduced functional DD-PC1 \cite{NVR.08} in self-consistent relativistic Hartree-Bogoliubov 
(RHB) calculations of energy surfaces (axial, triaxial, octupole), $\alpha$-decay energies and 
half-lives of SHN, in comparison to available data and previous theoretical studies.  
Section \ref{secII} introduces the general framework of
REDFs and, in particular, the functional DD-PC1 that will be used in illustrative calculations
throughout this work. The functional is tested in calculations of binding energies, 
ground-state quadrupole deformations, fission barriers, 
fission isomers, and $Q_{\alpha}$ values for even-even actinide nuclei. In Sec.~\ref{secIII} 
we apply the RHB framework based on the functional DD-PC1 and a separable pairing 
interaction in a description of triaxially deformed shapes and shape transitions of 
even-even superheavy nuclei. The microscopic, REDF-based,
quadrupole collective Hamiltonian model is used to study the effect of explicit 
treatment of collective correlations. $Q_\alpha$ values and half-lives for two chains 
of odd-even and odd-odd superheavy systems are  computed using a simple blocking 
approximation in axially symmetric self-consistent 
RHB calculations. Section \ref{secIV} summarizes the results and presents an outlook for future 
studies.
\section{\label{secII} The Relativistic Energy Density Functional DD-PC1}

Relativistic energy density functionals (REDF) provide an accurate, reliable, 
and consistent description of nuclear structure phenomena.
Semi-empirical functionals, adjusted to a microscopic nuclear matter equation 
of state and to bulk properties of finite nuclei, are applied to studies of
arbitrarily heavy nuclei, exotic nuclei far from stability, and even
systems at the nucleon drip-lines. REDF-based structure models
are being developed that go beyond the
mean-field approximation, and include collective correlations related to 
restoration of broken symmetries and to fluctuations of collective variables.

Although it originates in the effective interaction between nucleons, a generic density functional is not 
necessarily related to any given nucleon-nucleon potential and, in fact, some of the most successful modern 
functionals are entirely empirical. Until recently the
standard procedure of fine-tuning global nuclear density functionals was to perform a least-squares adjustment of 
a small set of free parameters simultaneously to empirical properties of symmetric and asymmetric nuclear matter, 
and to selected ground-state data of about ten spherical closed-shell nuclei. A new generation of semi-microscopic 
and fully microscopic functionals is currently being developed that will, on the one hand, establish a link with the 
underlying theory of strong interactions and, on the other hand, provide accurate predictions for 
a wealth of new data on short-lived nuclei far from stability. 
To obtain unique parameterizations, these functionals will have to be adjusted to a larger data set of ground-state properties, including both spherical and deformed nuclei \cite{Kor.10,Kor.12}. 

For a relativistic nuclear energy density functional the basic building blocks are  
densities and currents bilinear in the Dirac spinor field $\psi$ of the nucleon:
$\bar{\psi}\mathcal{O}_\tau \Gamma \psi\;$, $\mathcal{O}_\tau \in \{1,\tau_i\}$, 
   $\Gamma \in \{1,\gamma_\mu,\gamma_5,\gamma_5\gamma_\mu,\sigma_{\mu\nu}\}$.
$\tau_i$ are the isospin Pauli matrices and $\Gamma$ generically denotes the Dirac matrices. 
The isoscalar and isovector 
four-currents and scalar density are defined as expectation values of the 
corresponding operators in the nuclear ground state.
The nuclear ground-state is determined by the 
self-consistent solution of relativistic Kohn-Sham single-nucleon equations. 
To derive those equations it is useful to construct an 
interaction Lagrangian with four-fermion 
(contact) interaction terms in the various isospace-space channels: 
isoscalar-scalar   $(\bar\psi\psi)^2$, 
isoscalar-vector $(\bar\psi\gamma_\mu\psi)(\bar\psi\gamma^\mu\psi)$,
isovector-scalar $(\bar\psi\vec\tau\psi)\cdot(\bar\psi\vec\tau\psi)$,
isovector-vector $(\bar\psi\vec\tau\gamma_\mu\psi)
                         \cdot(\bar\psi\vec\tau\gamma^\mu\psi)$.
 A general Lagrangian can be written as a power series in the currents
 $\bar{\psi}\mathcal{O}_\tau\Gamma\psi$ and their derivatives, with 
higher-order terms representing in-medium many-body correlations. 

In Ref.~\cite{NVR.08} a Lagrangian was considered that includes
second-order interaction terms, with many-body correlations 
encoded in density-dependent strength functions.
A set of 10 constants, that control the strength and density dependence of 
the interaction Lagrangian, was fine-tuned 
in a multistep parameter fit exclusively to the experimental masses of 64 axially 
deformed nuclei in the regions $A\approx 150-180$ and $A\approx 230-250$. 
The resulting functional DD-PC1 has been further tested in calculations of 
binding energies, charge radii, deformation parameters, neutron skin 
thickness, and excitation energies of giant monopole and dipole resonances. 
The corresponding nuclear matter equation of state 
is characterized by the following properties at the saturation point: 
nucleon density $\rho_{sat}=0.152~\textnormal{fm}^{-3}$,
volume energy $a_v=-16.06$ MeV, surface energy $a_s=17.498$ MeV, 
symmetry energy $a_4 = 33$ MeV, and the nuclear matter compression modulus 
$K_{nm} = 230$ MeV. 

For a quantitative description of open-shell nuclei it is necessary to
consider also pairing correlations.
The relativistic Hartee-Bogoliubov  (RHB) framework \cite{VALR.05}
provides a unified description of particle-hole $(ph)$ and particle-particle 
$(pp)$ correlations on a mean-field level by combining two average
potentials: the self-consistent mean field that 
encloses all the long range \textit{ph} correlations, and a
pairing field $\hat{\Delta}$ which sums up the
\textit{pp}-correlations. In this work we perform axially-symmetric and triaxial 
calculations based on the RHB
framework with the \textit{ph} effective interaction derived from the DD-PC1 functional. 
A pairing force separable in momentum space: $\bra{k}V^{^1S_0}\ket{k'} = - G p(k) p(k')$ 
is used in the \textit{pp} channel. 
By assuming a simple Gaussian ansatz $p(k) = e^{-a^2k^2}$, the two parameters $G$ and $a$ 
were adjusted to reproduce the density dependence of the gap at the Fermi surface 
in nuclear matter, calculated with a Gogny force. For the D1S
parameterization~\cite{BGG.91} of the Gogny force the following
values were determined: $G=-728\;\mathrm{MeVfm}%
^{3}$ and $a=0.644\;\mathrm{fm}$ \cite{TMR.09a}. When
transformed from momentum to coordinate space, the force takes the form: 
\begin{equation}
V(\bff{r}_1,\bff{r}_2,\bff{r}_1^\prime,\bff{r}_2^\prime)
=G\delta \left(\bff{R}-\bff{R}^\prime \right)P(\bff{r})P(\bff{r}^\prime)
\frac{1}{2}\left(1-P^\sigma \right),
\label{pp-force}
\end{equation}
where $\bff{R}=\frac{1}{2}\left(\bff{r}_1+\bff{r}_2\right)$ and 
$\bff{r}=\bff{r}_1-\bff{r}_2$ denote the center-of-mass and the relative coordinates, 
and $P(\bff{r})$ is the Fourier transform of $p(k)$:
$P(\bff{r}) = 1/\left(4\pi a^2 \right)^{3/2}e^{-\bff{r}^2/4a^2}$.
The pairing force has finite range and, because of the presence of the factor
$\delta\left(  {\mbox{\boldmath $R$}}-{\mbox{\boldmath
$R$}}^{\prime}\right)  $, it preserves translational invariance. Even though
$\delta\left(  {\mbox{\boldmath $R$}}-{\mbox{\boldmath
$R$}}^{\prime}\right)  $ implies that this force is not completely separable
in coordinate space, the corresponding anti-symmetrized $pp$ matrix elements
can be represented as a sum of a finite number of separable terms in the basis
of a 3D harmonic oscillator \cite{NRV.10}. The force Eq.~(\ref{pp-force}) reproduces pairing 
properties of spherical and deformed nuclei calculated 
with the original Gogny force, but  with the important advantage that 
the computational cost is greatly reduced.

The Dirac-Hartree-Bogoliubov equations \cite{VALR.05} are solved by expanding the nucleon 
spinors in the basis of a 3D harmonic oscillator in Cartesian coordinates.
In this way both axial and triaxial nuclear shapes can be described. 
The map of the energy surface as a function of the quadrupole deformation is obtained
by imposing constraints on the axial and triaxial quadrupole moments. The method of 
quadratic constraint uses an unrestricted variation of the function
\begin{equation}
\label{eq:quadrupole-constraints1}
\langle \hat{H}\rangle +\sum_{\mu=0,2}{C_{2\mu}
   \left(\langle \hat{Q}_{2\mu}\rangle -q_{2\mu} \right)^2},
\end{equation}
where $\langle \hat{H}\rangle$ is the total energy, and $\langle \hat{Q}_{2\mu}\rangle$
denotes the expectation value of the mass quadrupole operators:
\begin{equation}
\label{eq:quadrupole-constraints2}
\hat{Q}_{20}=2z^2-x^2-y^2 \quad \textrm{and} \quad \hat{Q}_{22}=x^2-y^2 \;.
\end{equation}
$q_{2\mu}$ is the constrained value of the multipole moment, and $C_{2\mu}$
the corresponding stiffness constant.
\begin{figure}
\centering
\includegraphics[scale=0.60,angle=0]{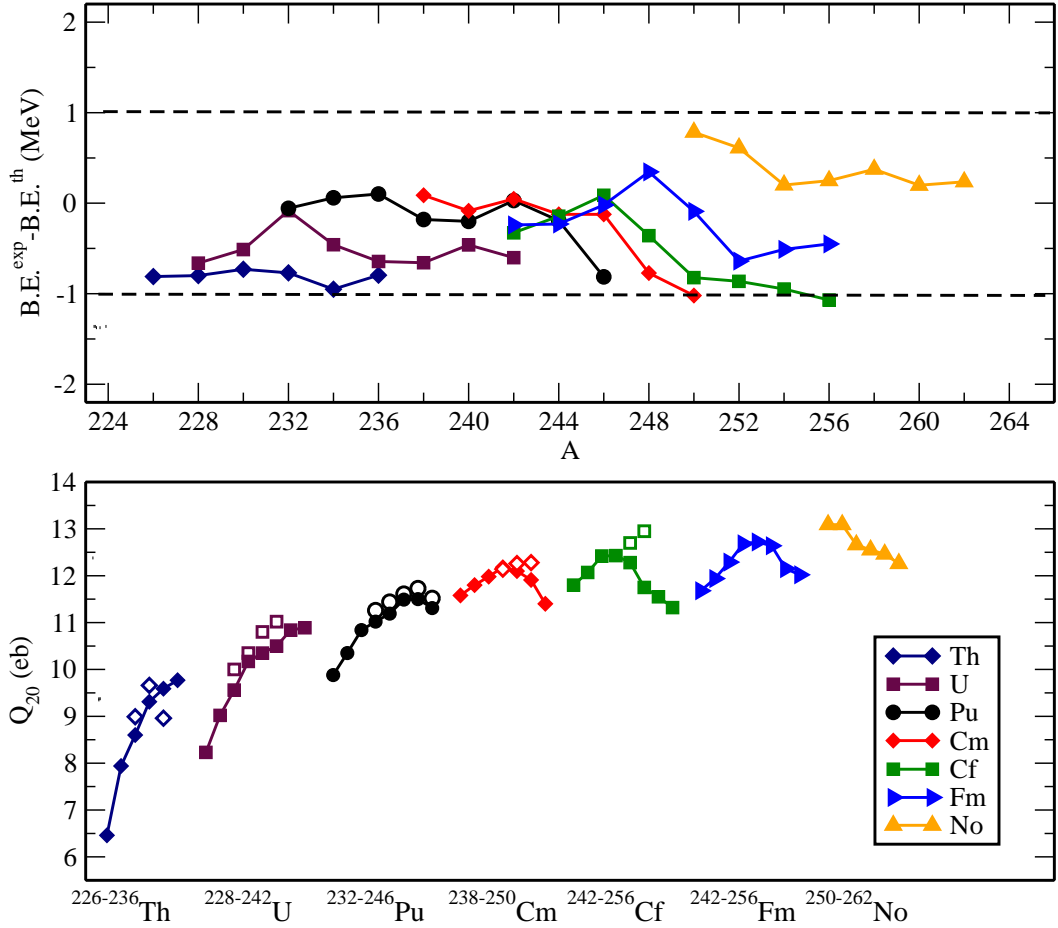}
\caption{(Color online) 
Absolute deviations of the self-consistent RHB ground-state binding energies from 
the experimental values \protect \cite{AW.03}, for the isotopic chains of Th, U, Pu, 
Cm, Cf, Fm, and No (upper panel). In the lower panel the calculated ground-state axial quadrupole 
moments are shown in comparison to data \protect \cite{RNT.01} (open symbols).} 
\label{masses&Q20}
\end{figure}

To illustrate the accuracy of the DD-PC1 functional in the calculation of ground-state 
properties of heavy nuclei, in Fig.~\ref{masses&Q20} we plot the 
results of self-consistent 3D RHB calculations for several 
isotopic chains in the actinide region: Th, U, Pu, Cm, Cf, Fm, and No. The 
deviations of the calculated binding energies from data \cite{AW.03}
show an excellent agreement between theory and experiment:  
the absolute difference between calculated and experimental binding energies 
is less than 1 MeV in all cases. An important results is also that the mass residuals do 
not display any notable dependence on the mass (neutron) number. 
The calculated ground-state quadrupole $Q_{20}$ moments 
are compared with available data \cite{RNT.01} in the lower panel of 
Fig.~\ref{masses&Q20}. One notices that the values predicted by the DD-PC1 functional reproduce 
in detail the isotopic trend of the empirical moments in the Th, U, Pu and Cm sequences, 
and are in very good agreement with the quadrupole moments of the Cf isotopes. 

The ``double-humped" fission barriers of actinide nuclei provide an important test for nuclear 
energy density functionals. In the review of self-consistent mean-field models
for nuclear structure \cite{BHR.03}, which also contains an extensive list of
references to previous studies of fission barriers using mean-field-based
models, Bender {\em et al.} compared paths in the deformation energy
landscape of $^{240}$Pu obtained with various Skyrme, Gogny and
relativistic mean-field (RMF)
interactions.  In general, relaxing constraints on symmetries lowers the fission
barriers. The predicted shapes are triaxial and reflection-symmetric at the first
barrier, and predominantly axial and reflection-asymmetric at the second barrier. 
The systematics of
axially symmetric fission barriers in Th, U, Pu, Cm and Cf nuclei, as well as for
superheavy elements $Z=108 - 120$, using several Skyrme and RMF mean-field
interactions, was investigated in Ref.~\cite{BBMR.04}. The fission barriers
of twenty-six even-Z nuclei with $Z = 90 - 102$, up to and beyond the
second saddle point, were calculated in Ref.~\cite{BQS.04} with the
constrained Hartree-Fock approach based on the Skyrme effective interaction
SkM$^*$. The fission barriers of $^{240}$Pu beyond the second
saddle point were also explored using the axially quadrupole constrained
RMF model with the PK1 effective interaction \cite{LGM.06}.
In a very recent optimization of the new Skyrme 
density functional UNEDF1 \cite{Kor.12}, excitation energies of fission isomers in $^{236,238}$U, 
$^{240}$Pu, and $^{242}$Cm, were added to the data set used to adjust the parameters of the 
functional. Compared to the original functional UNEDF0 \cite{Kor.10}, the inclusion of the new data 
allowed an improved description of fission properties of actinide nuclei. 
The effect of triaxial  deformation on fission barriers in the actinide region was recently also
explored in a systematic calculation of Ref.~\cite{Abu.10}, based on the RMF+BCS framework. 
The potential energy surfaces of actinide nuclei in the $(\beta_{20},\beta_{22},\beta_{30})$ deformation space 
(triaxial + octupole) were analyzed in the newest self-consistent mean-field plus BCS calculation based on 
relativistic energy density functionals \cite{Lu.12}. This study has shown the importance of the simultaneous 
treatment of triaxial and octupole shapes along the entire fission path. 

\begin{figure}[htpb]
\begin{center}
\scalebox{0.7}
{\includegraphics{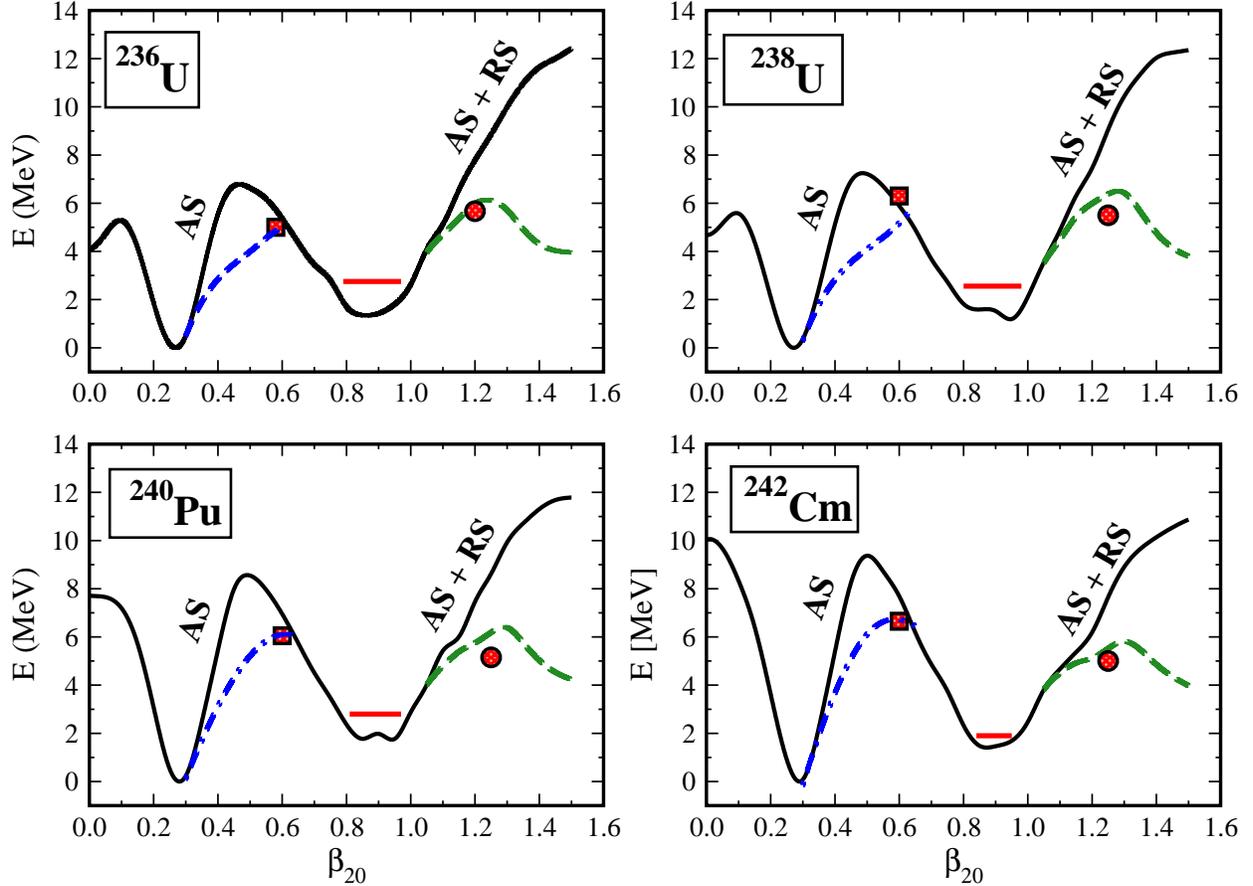}}
\caption{(Color online) Constrained energy curves of $^{236,238}$U, 
$^{240}$Pu, and $^{242}$Cm, as functions of the axial quadrupole deformation parameter.
Results of self-consistent axially and reflection-symmetric, triaxial, and axially reflection-asymmetric 
RMF+BCS calculations are denoted by solid (black), dot-dashed (blue), and dashed (green) curves, 
respectively. The red squares, lines, and circles denote the experimental 
values for the inner barrier height, the excitation energy of the fission isomer, and the height of 
the outer barrier, respectively. The data are from Ref.~\cite{Capote}. }
\label{actin_pes}
\end{center}
\end{figure}

The fission barriers calculated in the present work are shown in Fig. \ref{actin_pes}, 
where we plot the potential energy curves of $^{236,238}$U, 
$^{240}$Pu, and $^{242}$Cm, as functions of the axial quadrupole deformation parameter $\beta_{20}$. 
The deformation parameters are related to the multipole moments by the relation:
\be
\beta_{\lambda\mu} = {\frac{4\pi}{3A R^{\lambda}}}  \langle Q_{\lambda\mu} \rangle \; .
\ee
To be able to analyze the outer barrier heights considering reflection-asymmetric (octupole) shapes, the 
results displayed in this figure have been obtained in a self-consistent RMF plus BCS calculation that 
includes either triaxial shapes, or axially symmetric but reflection-asymmetric shapes.  
The interaction in the particle-hole channel is 
determined by the relativistic functional DD-PC1, and a density-independent $\delta$-force is the 
the effective interaction in the particle-particle channel. The pairing strength constants 
$V_n$ and $V_p$ are from Ref.~\cite{BMM.02}, where they were adjusted, together with the 
parameters of the relativistic functional PC-F1, to ground-state observables (binding energies, 
charge and diffraction radii, surface thickness and pairing gaps) of spherical nuclei, with pairing 
correlations treated in the BCS approximation. In many cases the functionals PC-F1 \cite{BMM.02} 
and DD-PC1 predict similar results for ground state properties (cf. for instance, Ref.~\cite{Li.10}), and 
reproduce the empirical pairing gaps. Thus we assume that, without any further adjustment, 
the same pairing strength parameters can be used in RMF+BCS calculations with the functional 
DD-PC1.

The solid (black) curves correspond to binding energies calculated with the constraint on the 
axial quadrupole moment, assuming axial and reflection symmetry. The absolute minima of these curves 
determine the energy scale (zero energy). The dot-dashed (blue) curves denote paths of minimal energy  
in calculations that break axial symmetry with constraints on quadrupole axial $Q_{20}$ and triaxial 
$Q_{22}$ moments. Finally, the dashed (green) curves are paths of minimal energy obtained in 
axially symmetric calculations that break reflection symmetry (constraints on the quadrupole moment 
$Q_{20}$ and the octupole moment $Q_{30}$). The red squares, lines, and circles denote the experimental 
values for the inner barrier height, the excitation energy of the fission isomer, and the height of the outer barrier, 
respectively. The data are from Ref.~\cite{Capote}. 
\begin{figure}[htpb]
\begin{center}
\scalebox{0.55}
{\includegraphics{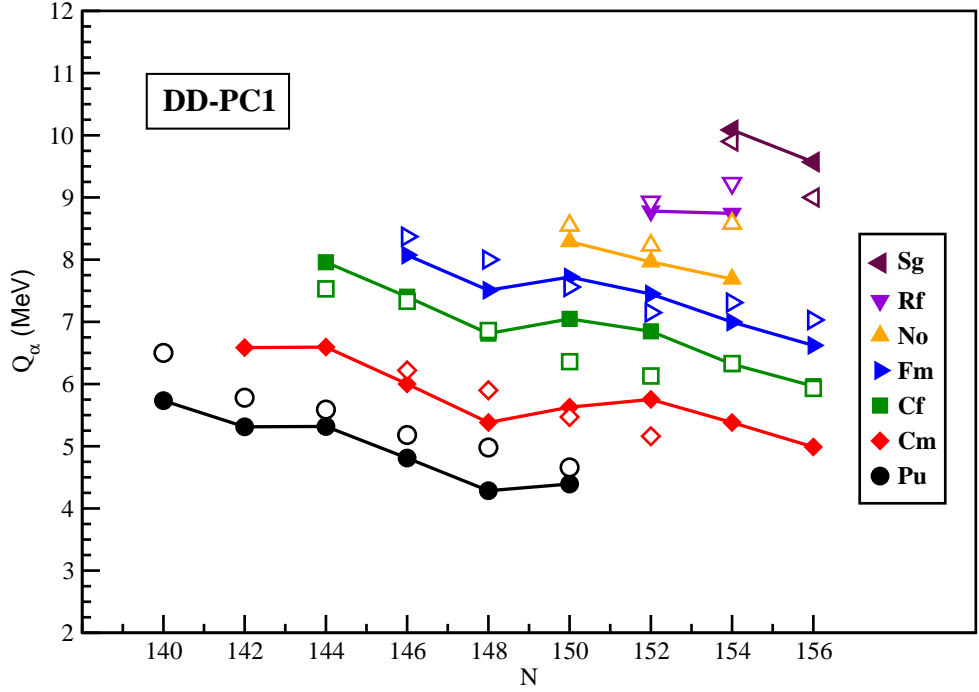}}
\caption{(Color online) $Q_{\alpha}$ values for even-even actinide chains obtained in a self-consistent 
axially symmetric RHB calculations using the functional DD-PC1 and the separable pairing 
interaction Eq.~(\ref{pp-force}). The theoretical values (filled symbols) are connected by lines 
and compared to data (open symbols) \cite{AW.03}.}
\label{actin}
\end{center}
\end{figure}

The excitation energies of fission isomers are fairly well reproduced by the axially symmetric and reflection 
symmetric calculation, but the paths constrained by these symmetries overestimate the height of the 
inner and outer barriers. The inclusion of triaxial shapes lowers the inner barrier by $\approx 2$ MeV, 
that is, the axially symmetric barriers in the region $\beta_{20} \approx 0.5$ are bypassed through the 
triaxial region, bringing the height of the barriers much closer to the empirical values. As shown in the 
figure, the inclusion of octupole shapes (axial, reflection-asymmetric calculations) is essential to reproduce 
the height of the outer barrier in actinide nuclei. A very good agreement with data is obtained by following 
paths through shapes with non-vanishing octupole moments. With the present implementation of the model 
we could not simultaneously calculate both octupole and triaxial shapes, but such a study was recently 
performed in the RMF+BCS framework by Bing-Nan Lu and collaborators \cite{Lu.12}. It was shown 
that not only the inner barrier, but also the reflection-asymmetric outer barrier is lowered by the inclusion 
of triaxial deformations. This effect is of the order of 0.5 - 1 MeV, and it accounts for 10\% to 20\% of the 
barrier height. Considering that the functional DD-PC1 was adjusted only to the binding energies of 
the absolute axial minima (masses) of deformed nuclei, the results shown in Fig. \ref{actin_pes} 
reproduce the experimental values surprisingly well and, therefore, 
appear to be very promising for its extrapolation to the region of superheavy nuclei. We note here 
that, except for the results of Fig. \ref{actin_pes}, all the other calculations reported in this 
work have been performed in the RHB framework using the functional DD-PC1 and the separable pairing 
interaction Eq.~(\ref{pp-force}).

Figure \ref{actin} illustrates the accuracy of the functional DD-PC1 in the axially symmetric RHB calculation 
of $Q_{\alpha}$ values, that is, energies of $\alpha$ particles emitted by even-even actinide nuclei.  
The calculated values are plotted in comparison to data \cite{AW.03}. Even in this simple calculation that 
assumes axial symmetry, the model reproduces the empirical trend of $Q_{\alpha}$ values. 
The few cases for which we find a somewhat larger deviation from data most probably point to 
a more complex potential energy surface, possibly including shape coexistence. It is interesting to 
note that on the quantitative level the theoretical results are very similar to those 
obtained in the self-consistent non-relativistic Hartree-Fock-Bogoliubov calculation based on the 
Skyrme functional SLy4 \cite{HN.02,CHN.05}. 

Summarizing this section, it has been shown that self-consistent mean-field calculations based on the relativistic 
energy density functional DD-PC1 predict binding energies, ground-state quadrupole deformations, fission barriers, 
fission isomers, and $Q_{\alpha}$ values for even-even actinide nuclei in very good agreement with data. In the 
next section we apply the RHB framework based on the functional DD-PC1 and the separable pairing 
interaction Eq.~(\ref{pp-force}) to an illustrative study of deformed shapes and shape transitions of 
superheavy nuclei.

\section{\label{secIII} Shape transitions in superheavy nuclei}
In a very recent study of fission barriers and fission paths in even-even superheavy 
nuclei with $Z= 112-120$ \cite{Abu.12}, DD-PC1 was used, together with two other relativistic 
energy density functionals, in a systematic RMF+BCS calculation of potential energy 
surfaces, including triaxial and octupole shapes. It was shown that low-Z and low-N 
nuclei in this region are characterized by axially symmetric inner fission barriers. 
With the increase of Z and/or N, in some of these nuclei several competing fission 
paths appear in the region of the inner barrier. Allowing for triaxial shapes lowers the outer 
fission barrier by $1.5 - 3$ MeV, and in many nuclei the lowering induced by triaxiality is even more 
important than the one due to octupole deformation. 

\begin{figure}[htpb]
\begin{center}
\scalebox{0.6}
{\includegraphics{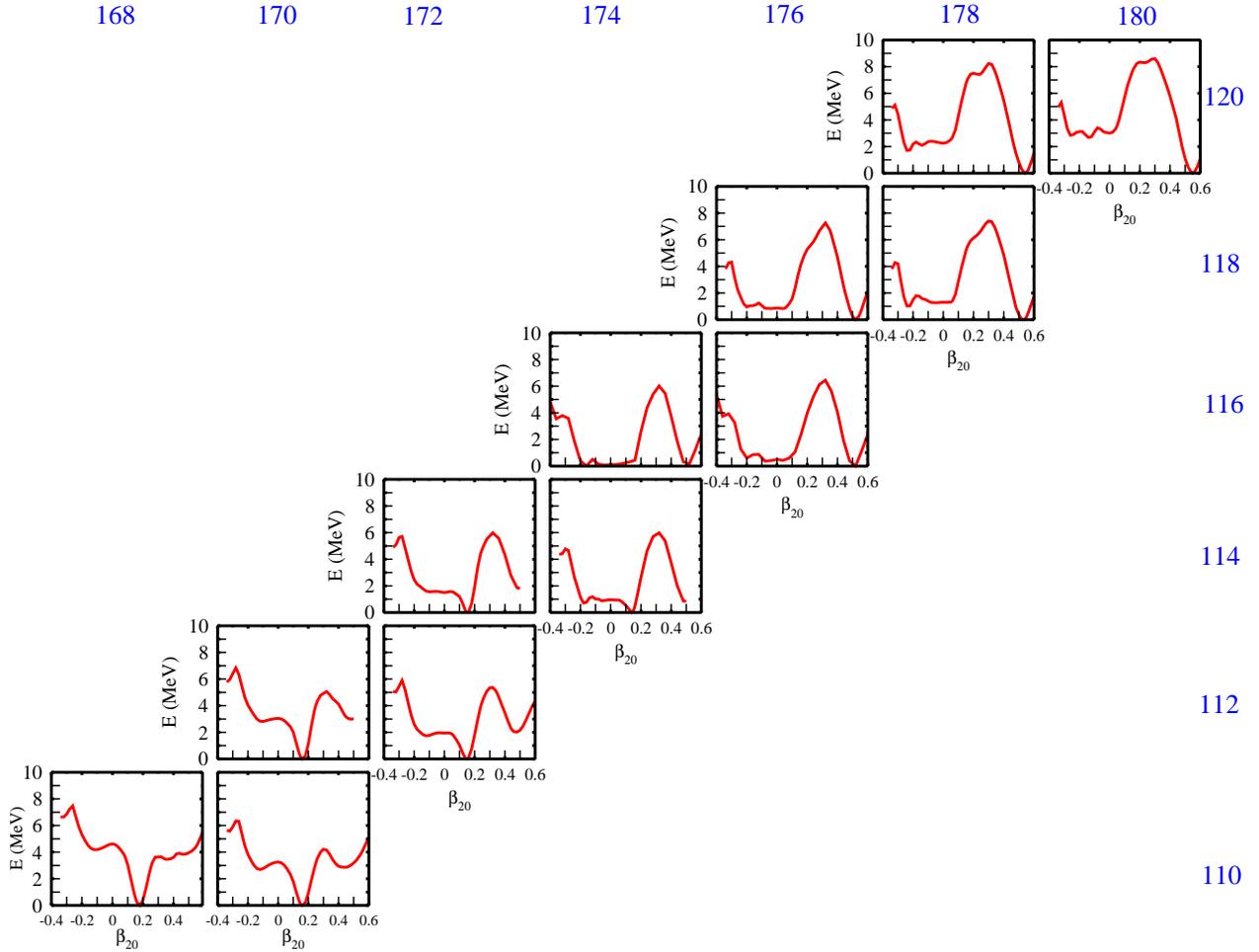}}
\caption{(Color online) Self-consistent RHB axially symmetric energy curves of isotopes 
in the $\alpha$-decay chains of  $^{298}$120 and $^{300}$120, as functions of the quadrupole deformation parameter.}
\label{pes-table}
\end{center}
\end{figure}

The variation of ground-state shapes is governed by the evolution of the shell structure 
of single-nucleon orbitals. In very heavy deformed nuclei, in particular, the density 
of single-nucleon states close to the Fermi level is rather large, and even small variations  
in the shell structure predicted by different effective interactions can lead to markedly distinct 
equilibrium deformations. To illustrate the rapid change of equilibrium shapes for the heaviest nuclear systems, 
Figure \ref{pes-table} displays the results of self-consistent axially symmetric RHB calculations of 
isotopes in the $\alpha$-decay chains of  two superheavy nuclei: $^{298}$120 and $^{300}$120. 
The quadrupole energy curves are plotted as functions of the deformation parameter $\beta_{20}$.  
Lighter systems around $Z=110$ are characterized by well developed prolate minima around 
$\beta_{20} \approx 0.2$, whereas intermediate nuclei display both oblate and prolate minima 
at small deformation, and the heaviest isotopes appear to be slightly oblate. 
Another characteristic of the energy curves is the shift of the saddle point to smaller deformations 
with the increase in the mass number, while the barriers become wider.
Since the prolate and 
oblate minima can be connected through triaxial shapes without a barrier, these energy curves 
show the importance of performing more realistic calculations, including triaxial, and for large 
deformations, octupole shapes.  

\begin{figure}
\includegraphics[scale=0.375]{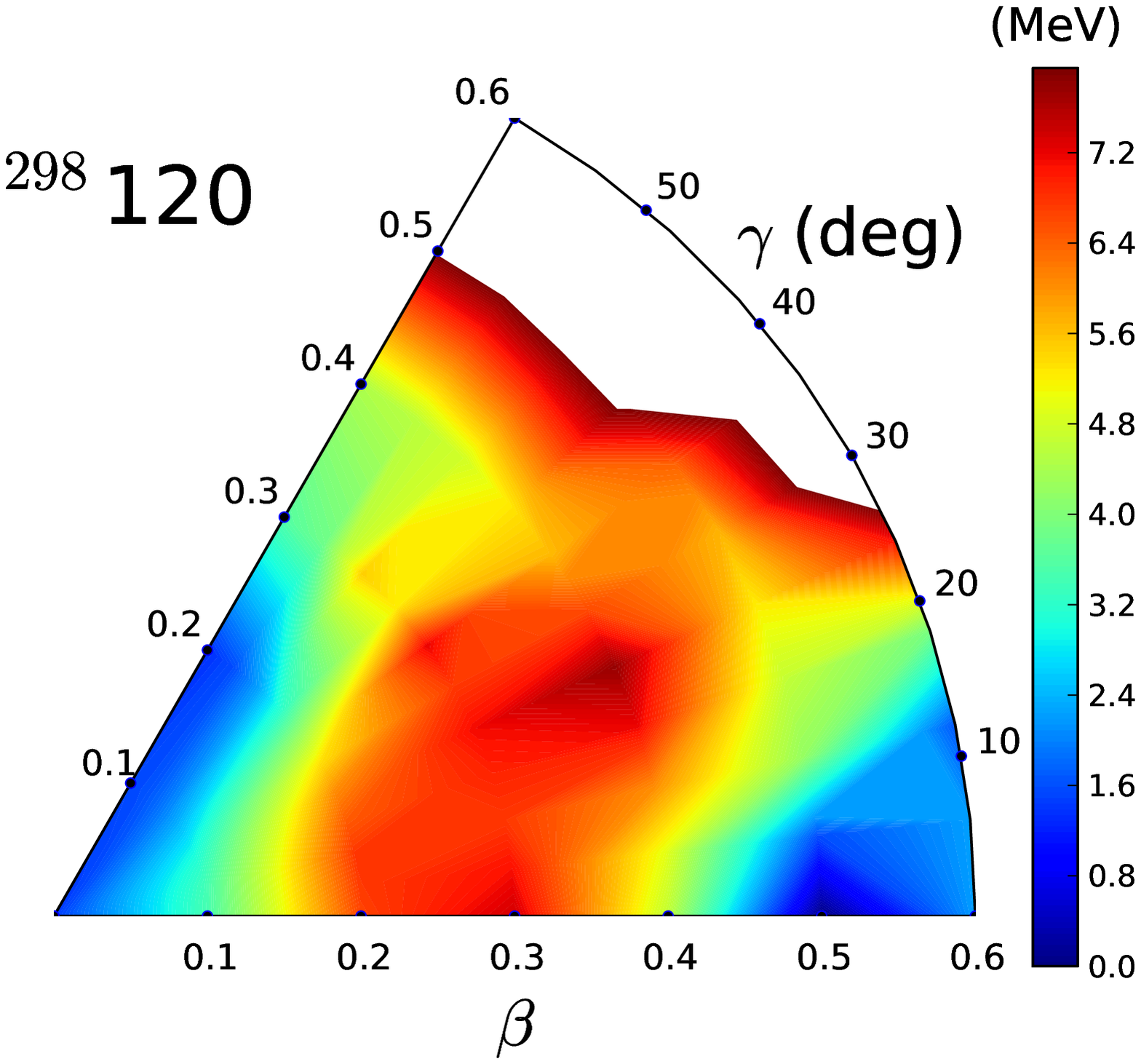}
\includegraphics[scale=0.375]{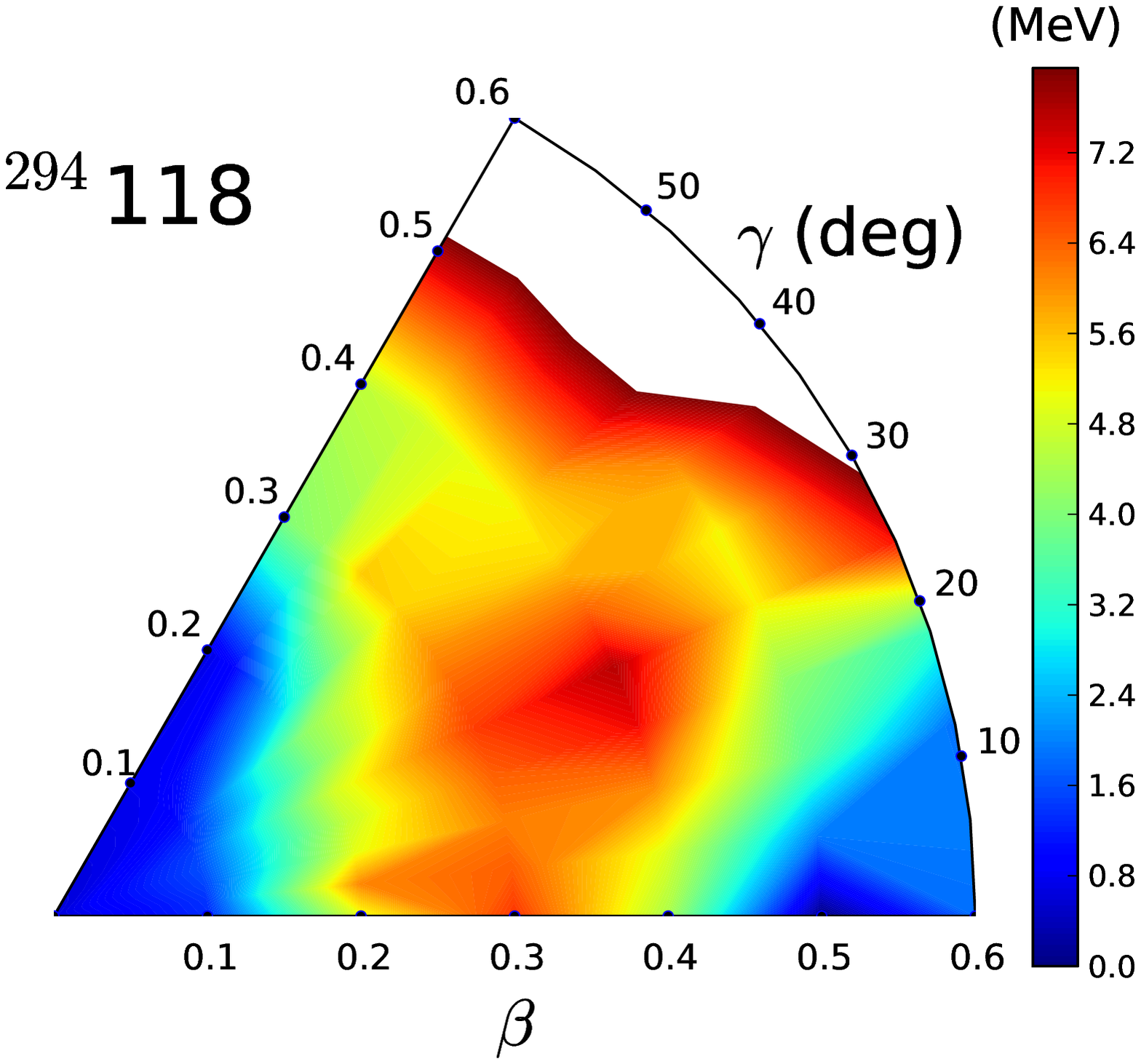}\\
\includegraphics[scale=0.375]{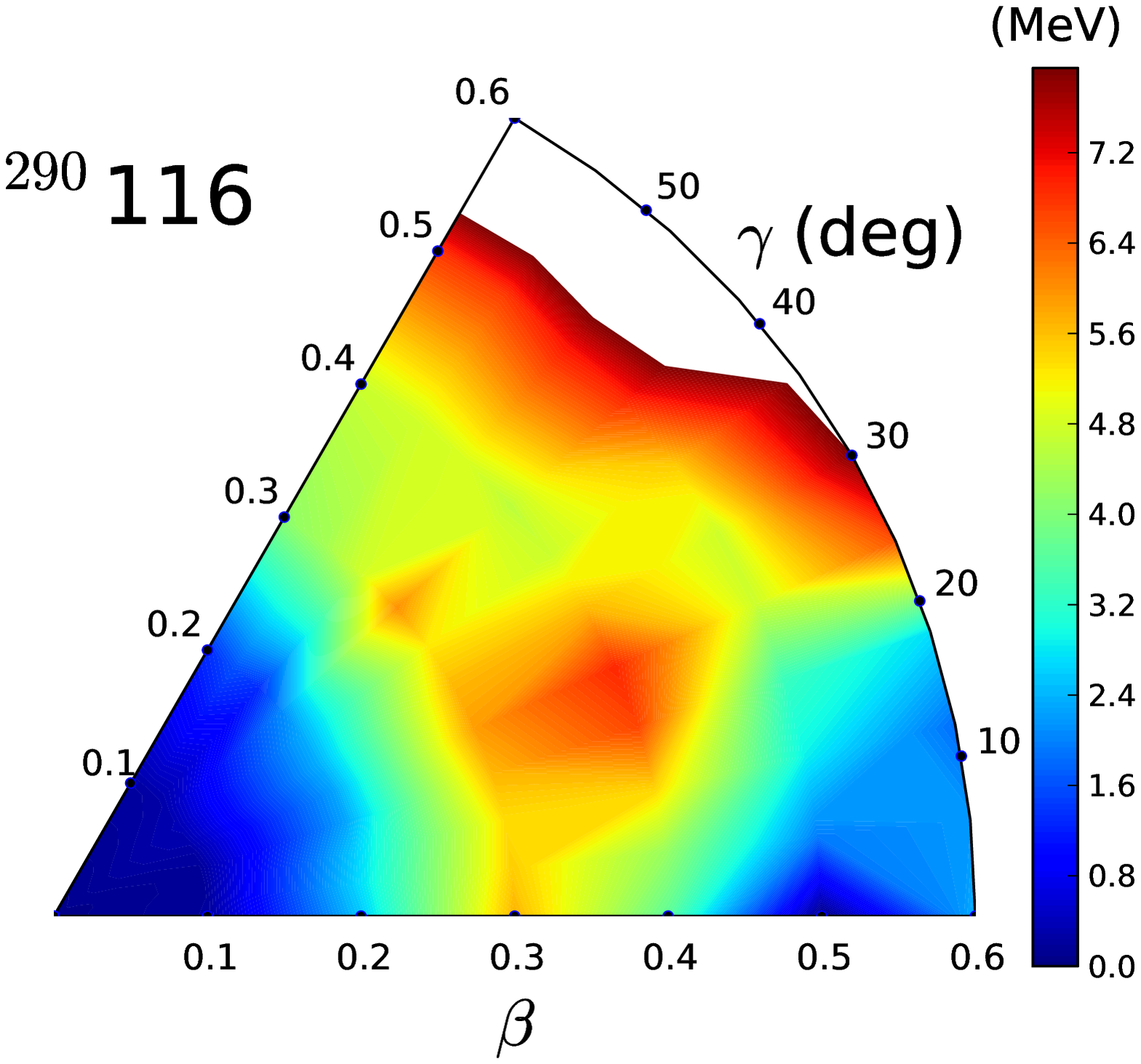}
\includegraphics[scale=0.375]{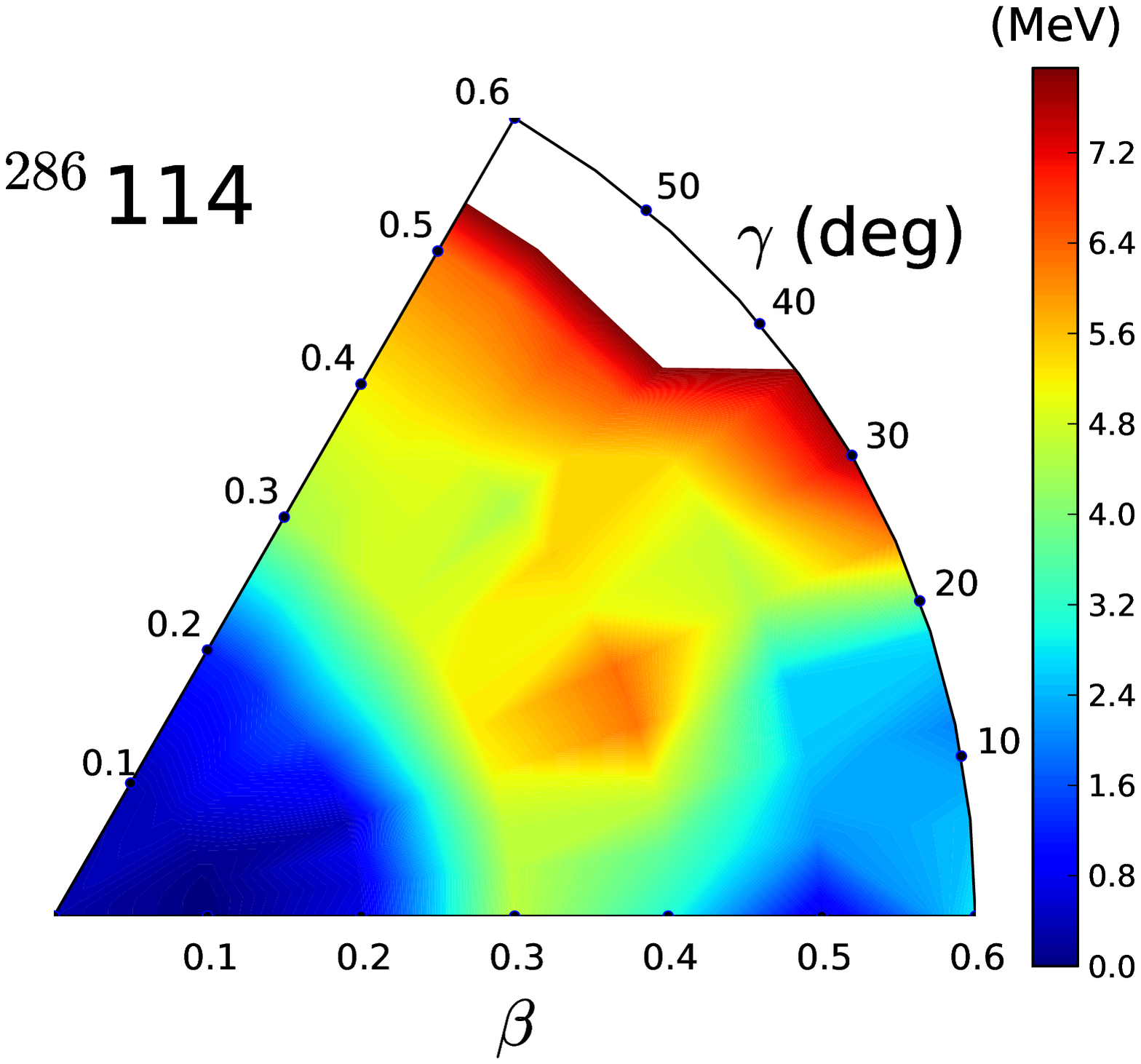}\\
\includegraphics[scale=0.375]{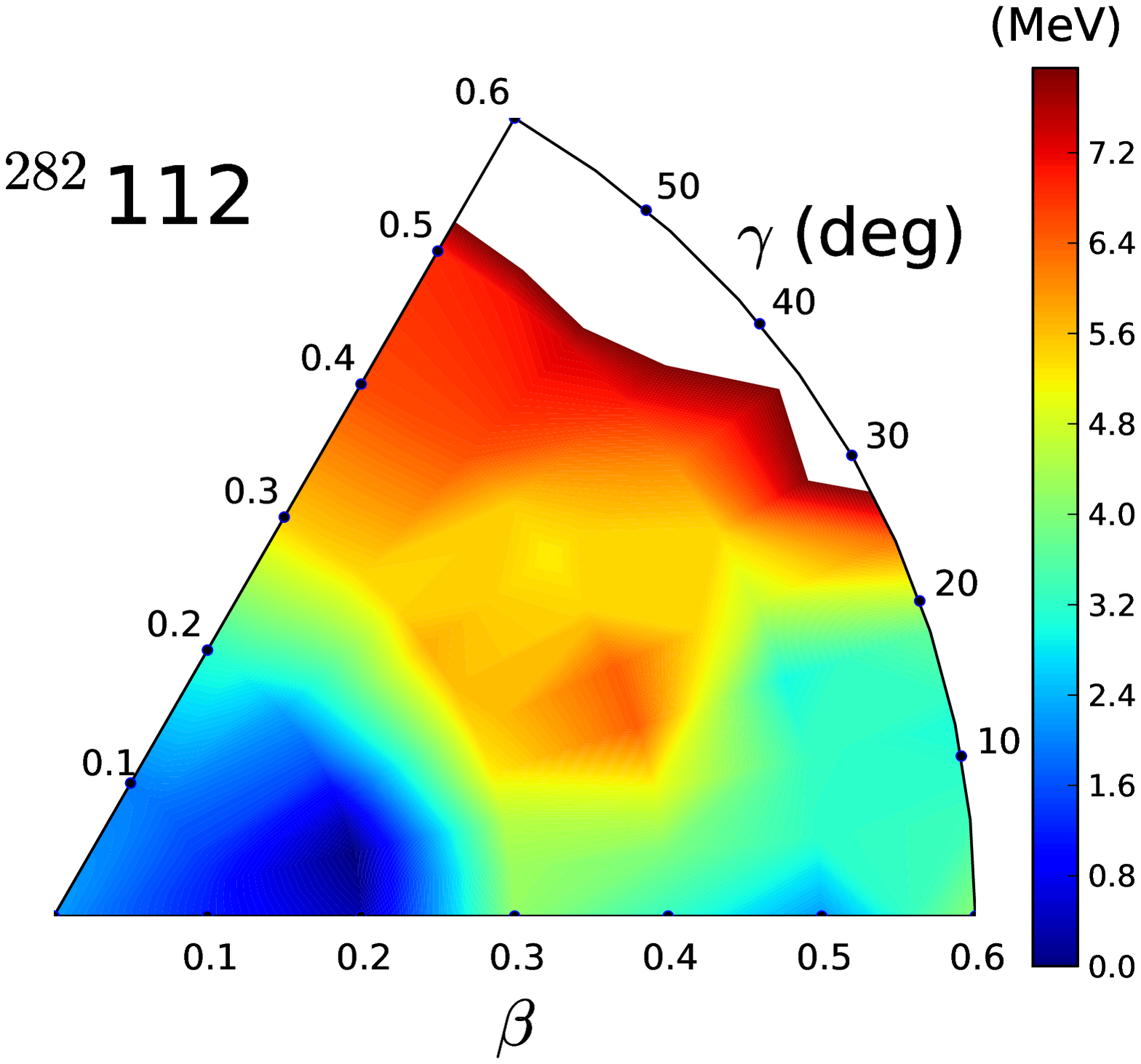}
\includegraphics[scale=0.375]{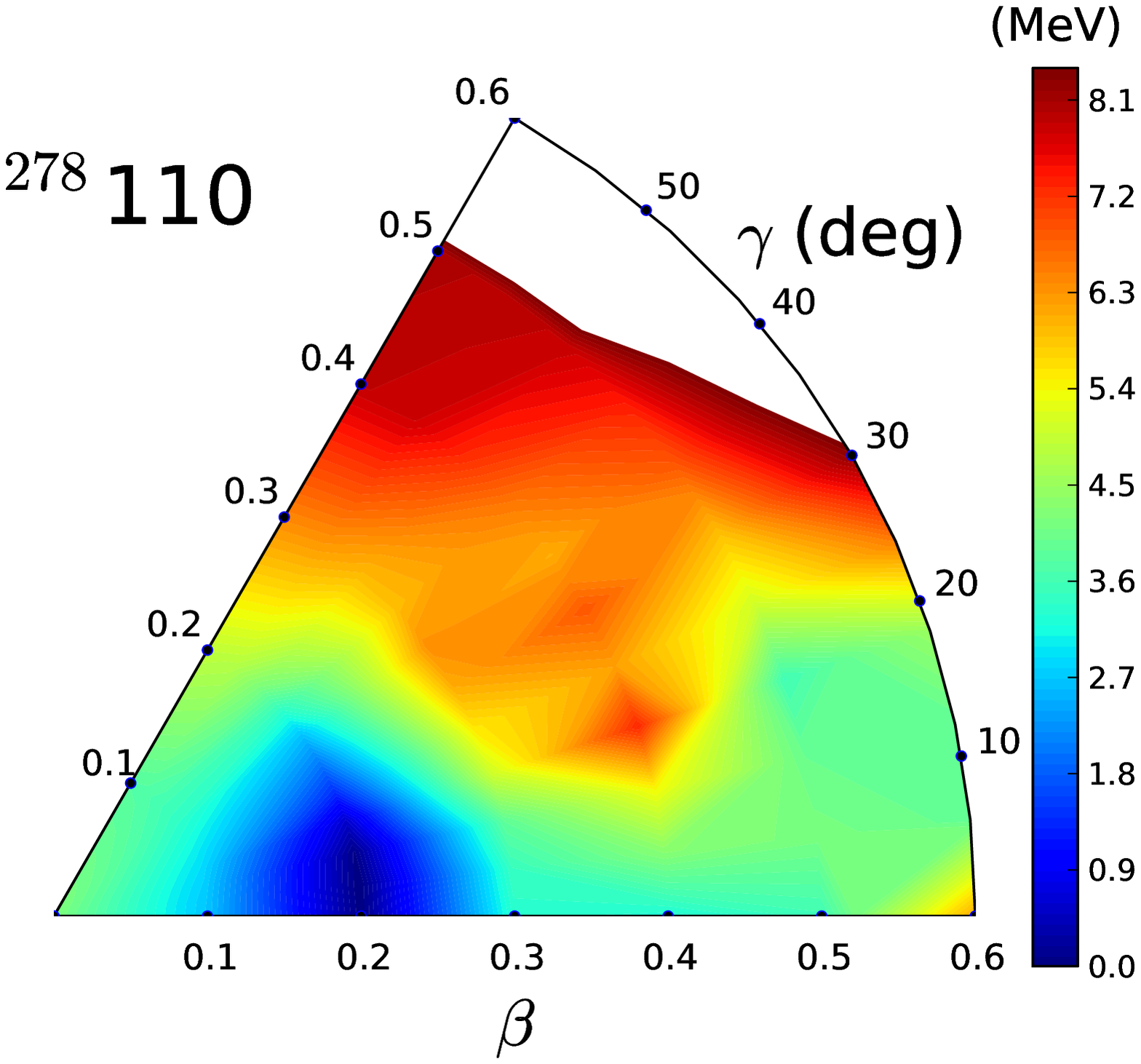}
\caption{(Color online) Self-consistent RHB triaxial energy maps of the
even-even isotopes in the $\alpha$-decay chain of $^{298}$120 in the
$\beta-\gamma$ plane ($0\le \gamma \le 60^\circ$). Energies are
normalized  with respect to the binding energy of the absolute minimum. }
\label{pes1}
\end{figure}

\begin{figure}
\includegraphics[scale=0.375]{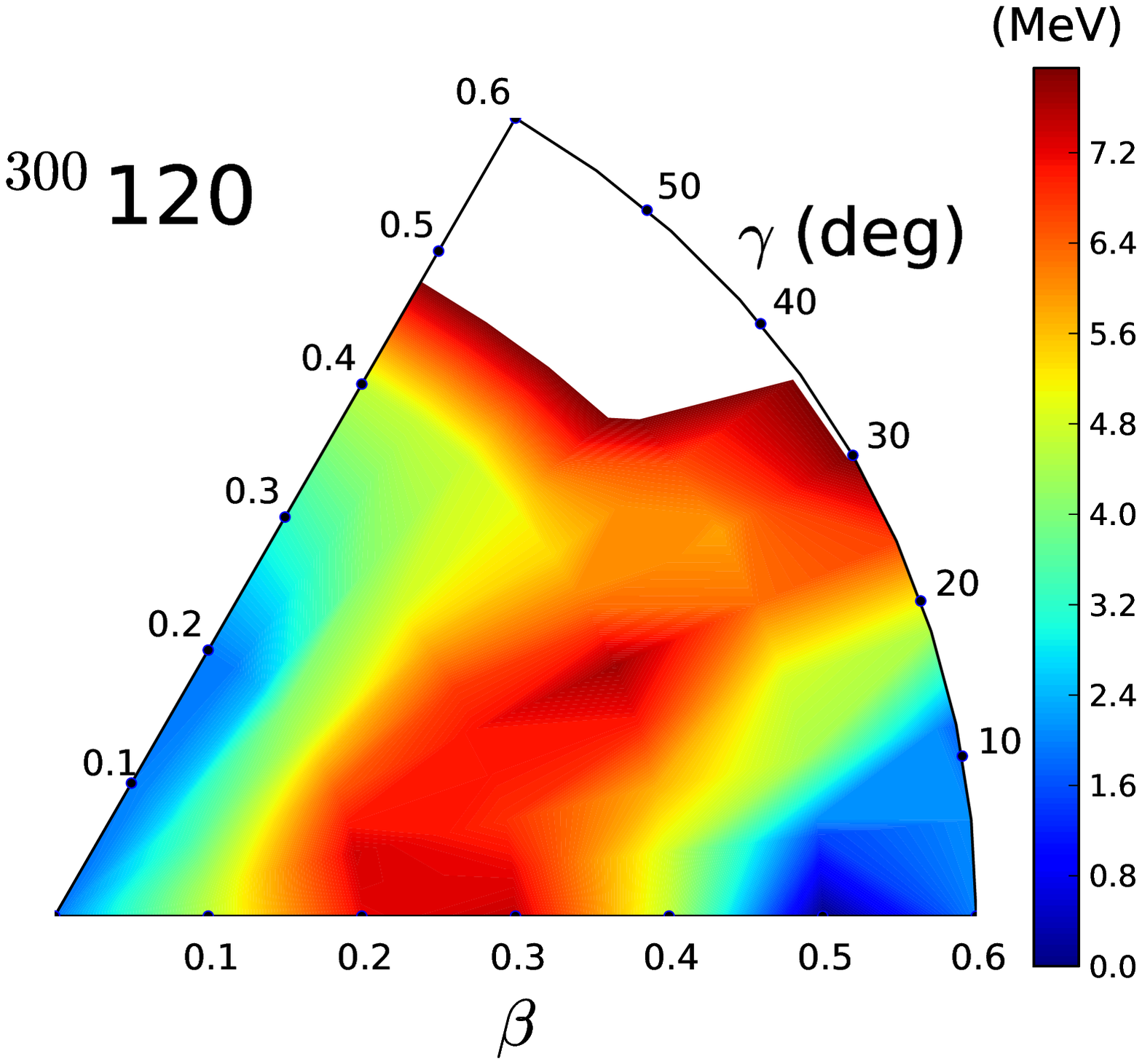}
\includegraphics[scale=0.375]{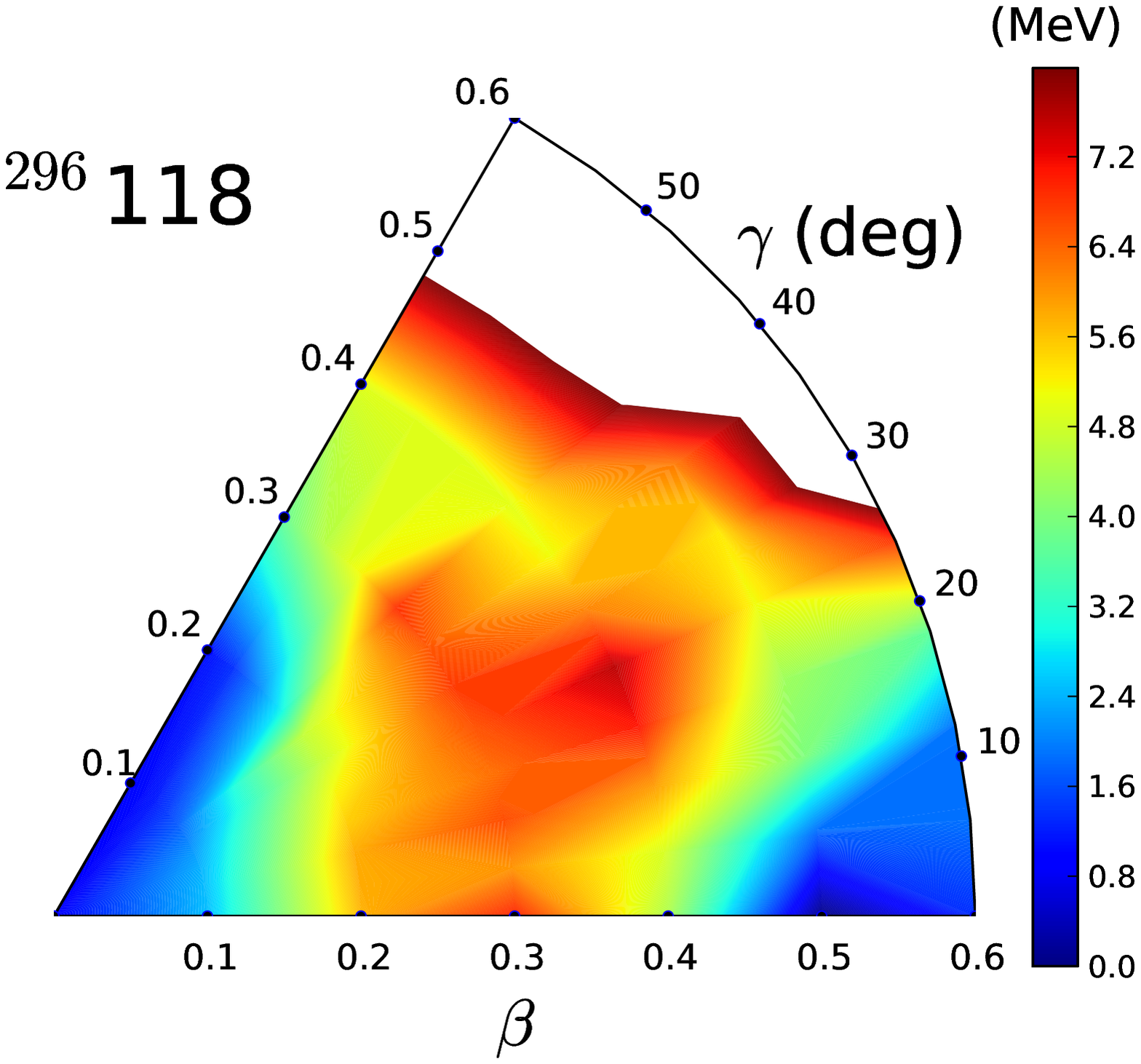}\\
\includegraphics[scale=0.375]{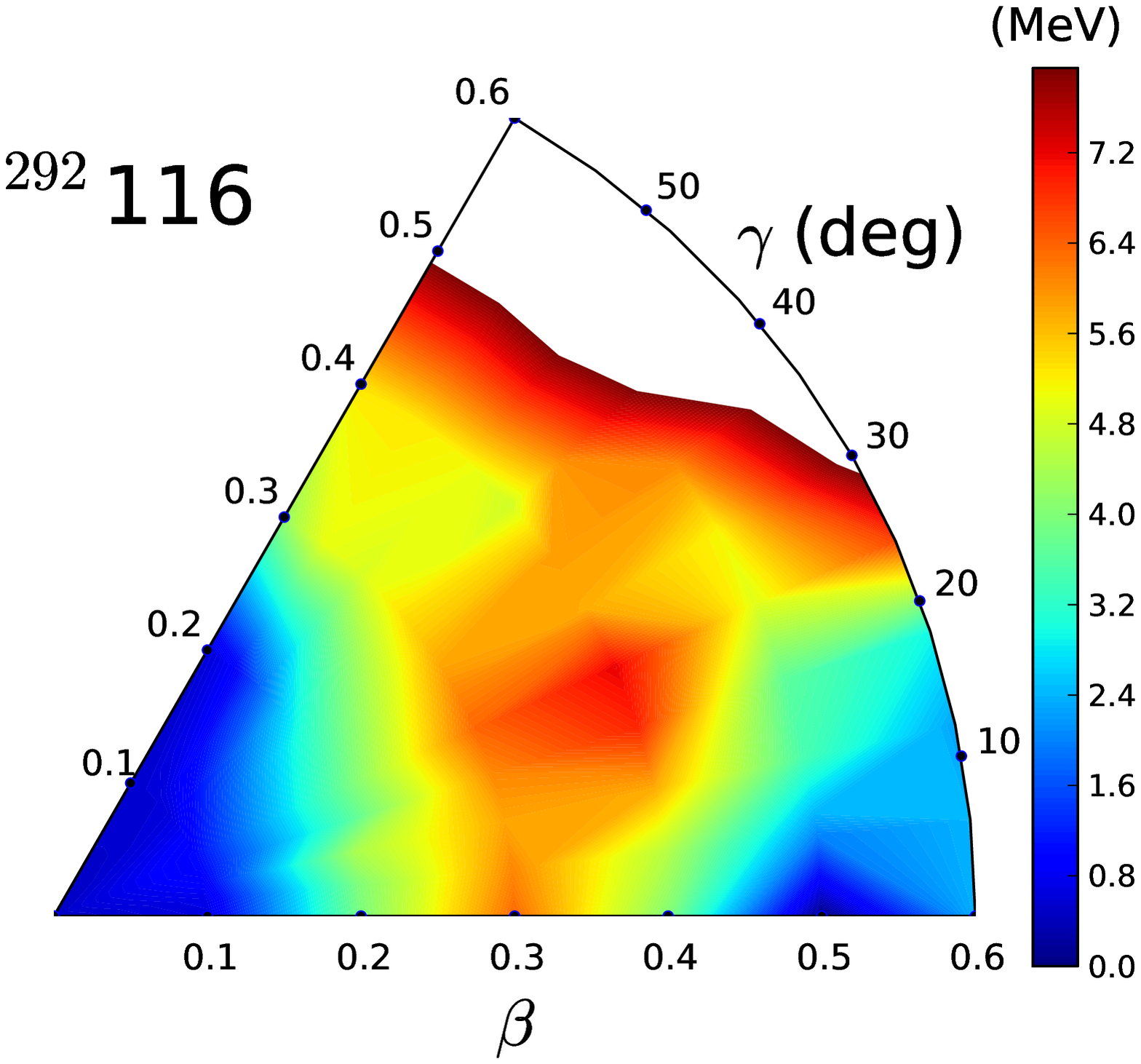}
\includegraphics[scale=0.375]{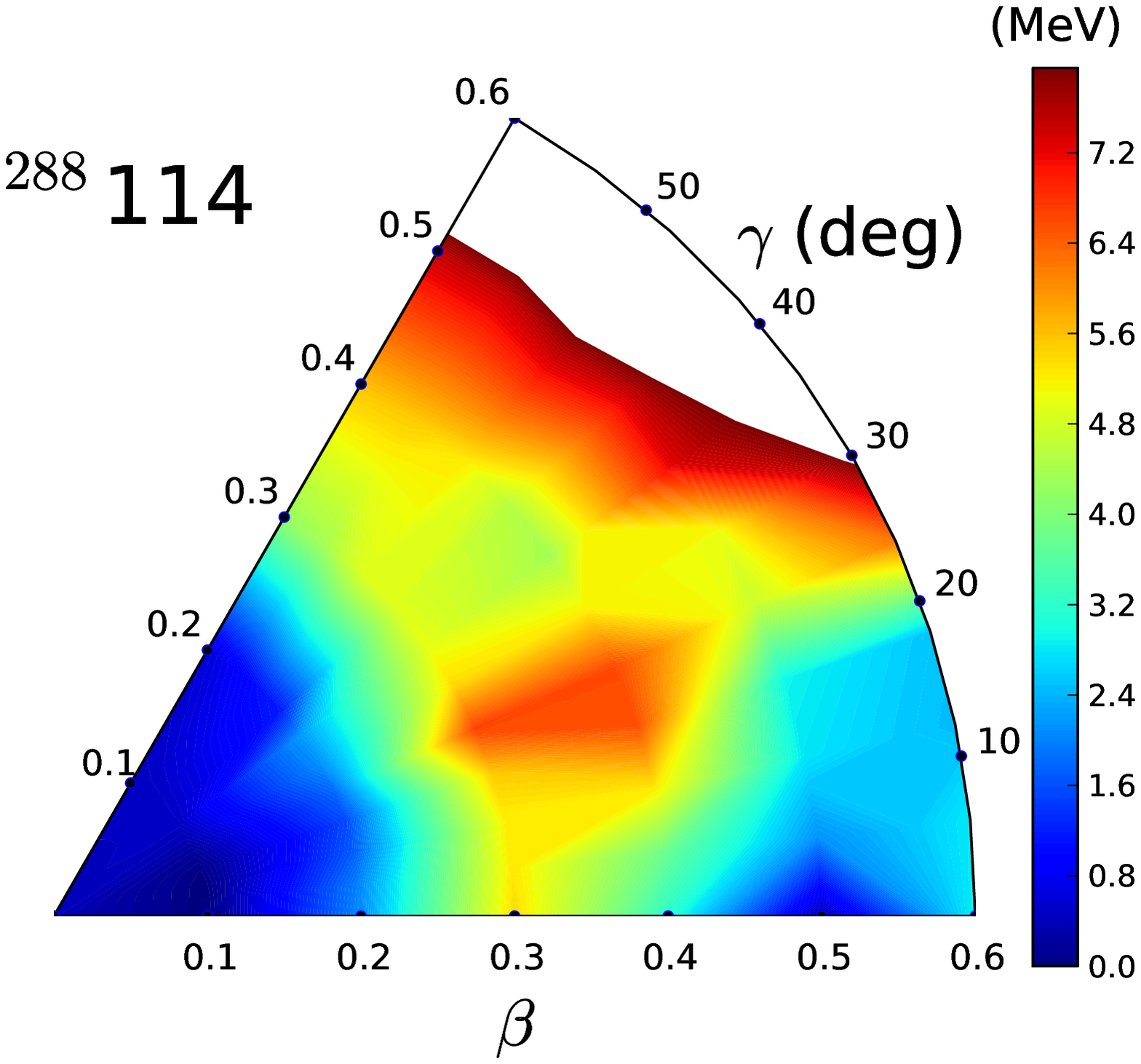}\\
\includegraphics[scale=0.375]{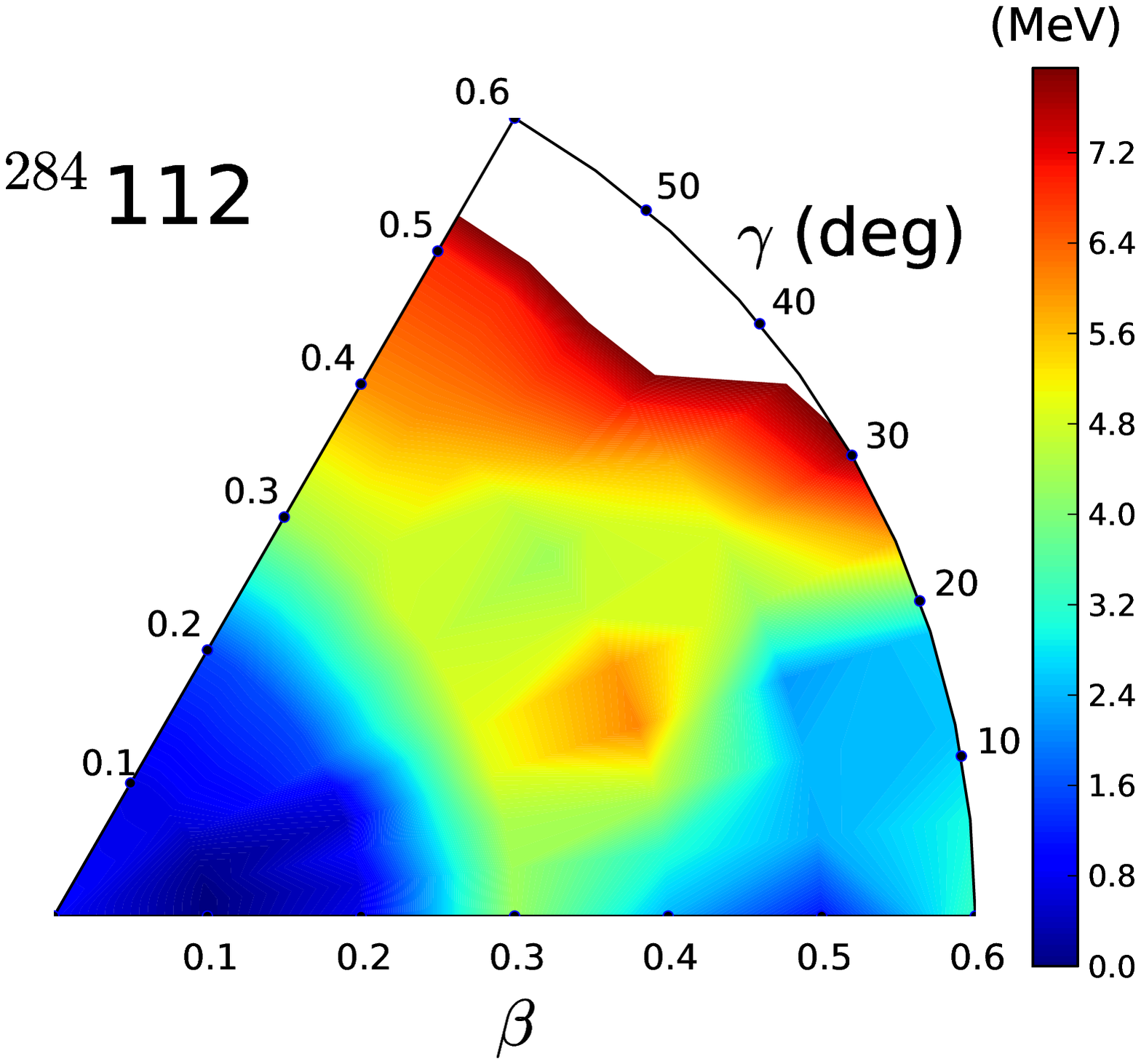}
\includegraphics[scale=0.375]{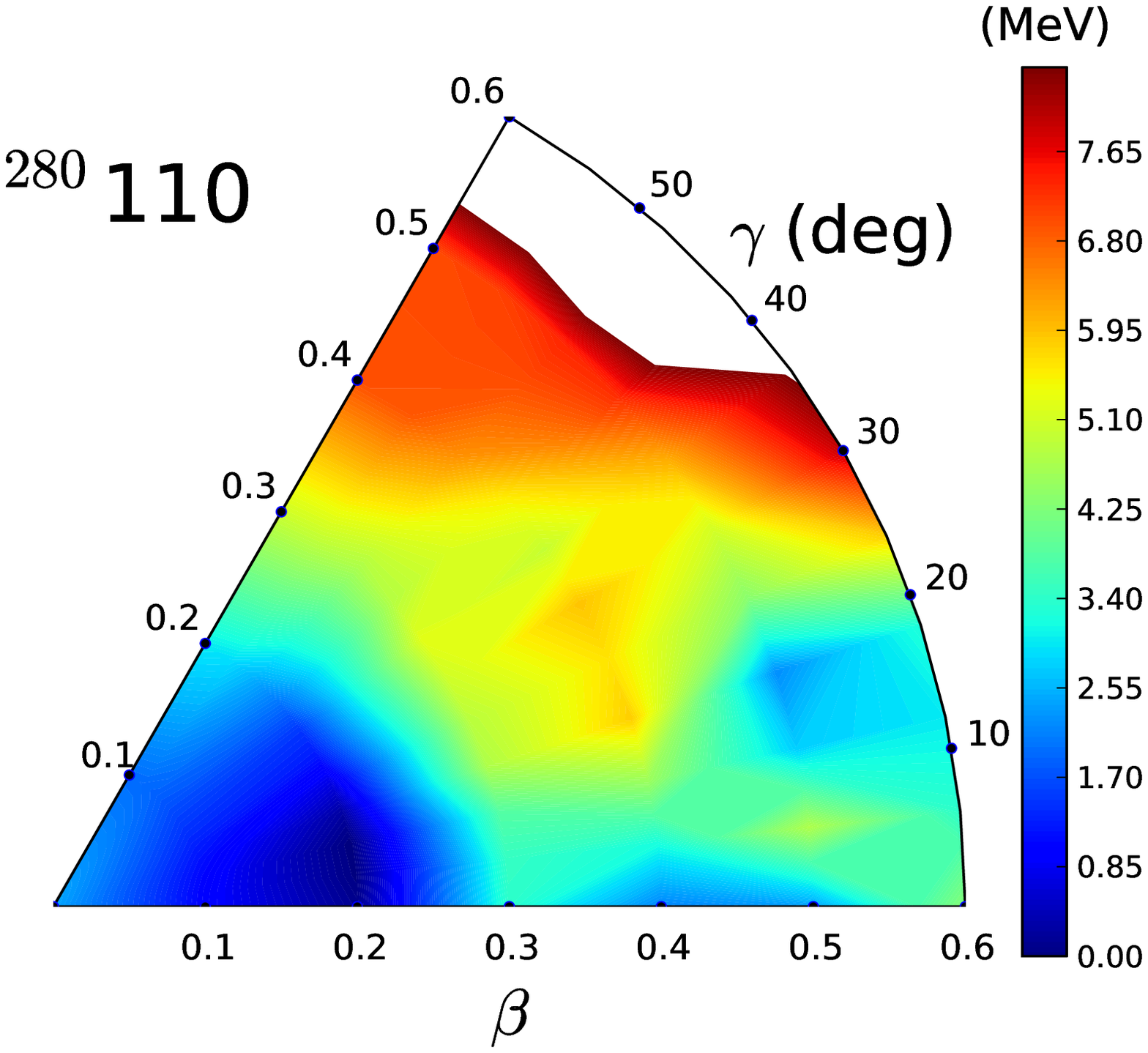}
\caption{(Color online) Same as
described in the caption to Fig. \ref{pes1}
but for the the $\alpha$-decay chain of $^{300}$120.
 }
\label{pes2}
\end{figure}

This is shown in Figs. \ref{pes1} and \ref{pes2}, where we plot the corresponding triaxial RHB energy 
surfaces in the $\beta-\gamma$ plane ($0\le \gamma \le 60^\circ$) for isotopes in the $\alpha$-decay 
chains of $^{298}$120 and $^{300}$120, respectively. In 
both chains the heaviest systems display soft oblate axial shapes with minima that extend from the spherical 
configuration to $|\beta_{20}| \approx 0.4$ ($Z=120$) and  $|\beta_{20}| \approx 0.3$ ($Z=118$). We do not 
have to consider the deep prolate minima at $\beta_{20} > 0.5$ because the inclusion of reflection asymmetric 
shape degrees of freedom (octupole deformation) drastically reduces or removes completely the outer barrier. 
A low outer barrier implies a high probability for spontaneous fission, such that the prolate superdeformed 
states are not stable against fission\cite{Moller2009}. 
In contrast to the actinides shown in Fig.~\ref{actin_pes}, superheavy
nuclei are actually characterized by  a ``single-humped" fission barrier. 
As already noted, in the present implementation of the model triaxial and octupole deformations cannot be 
taken into account simultaneously. Reflection asymmetric shape degrees of freedom, however, play no role 
at small and moderate deformations that characterize ground-state configurations of the superheavy systems 
considered here. The intermediate nuclei with $Z=116$  are essentially spherical but soft both in 
$\beta$ and $\gamma$, whereas prolate deformed mean-field minima develop in the lighter systems with 
$Z=114$, $Z=112$ and $Z=110$. The predicted evolution of shapes is consistent with results obtained using 
the self-consistent Hartree-Fock-Bogoliubov framework based on Skyrme 
functionals \cite{cwiok96,cwiok99,CHN.05}. 

\begin{figure}
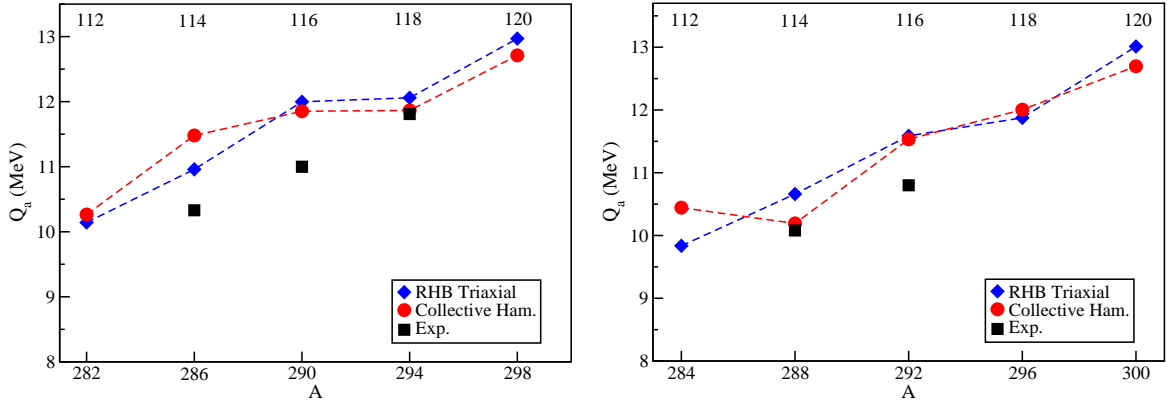

\includegraphics[scale=0.315]{Qa_116_290.eps}\quad
\includegraphics[scale=0.315]{Qa_116_292.eps}
\caption{(Color online) $Q_\alpha$ values for the $\alpha$-decay chains of $^{298}$120 (left) and 
$^{300}$120 (right). The theoretical values are calculated as the difference between the mean-field 
minima of the parent and daughter nuclei (blue diamonds), and as the difference between the 
energies of the $0^+$ ground states of the quadrupole collective Hamiltonian (red circles). 
The data (squares) are from Ref.~\cite{og2007}.}
\label{qeven}
\end{figure}

The two main decay modes in this region are $\alpha$-emission and spontaneous fission. 
The theoretical $\alpha$-decay energies, denoted by (blue) diamonds in Fig.~\ref{qeven}, are calculated 
as the difference between the absolute minima of the energy maps of the parent and daughter nuclei, 
shown in Figs.~\ref{pes1} and \ref{pes2}. For the mean-field ground state we take the minimum with the 
highest barrier with respect to fission, that is, we do not consider superdeformed minima with very low fission 
barriers. The theoretical $Q_\alpha$ values are shown in comparison to available data for 
the $\alpha$ decay properties of superheavy nuclei \cite{og2007}. 
The trend of the data is obviously reproduced by the calculation, and the largest difference between 
theoretical and experimental values is less than 1 MeV. This is a rather good result, considering that 
equilibrium nuclear shapes change rapidly in the two $\alpha$-decay chains 
and that the calculation is performed on the mean-field level. In general the level of 
agreement with experiment is similar to the one found in the case of actinide nuclei (cf. Fig.~\ref{actin}), 
but not quite the same as that obtained using the macro-micro model specially adapted to heaviest nuclei 
(HN) \cite{SobPom.07,MPS.01,Sob.10,Sob.11}. It is interesting, though, that our prediction 
both for the $\alpha$-decay energy and half-life of the $A=298$ isotope of the new element $Z=120$ 
are consistent with those of the HN macro-micro model \cite{Sob.11}.

%
%
%
%
\begin{figure}
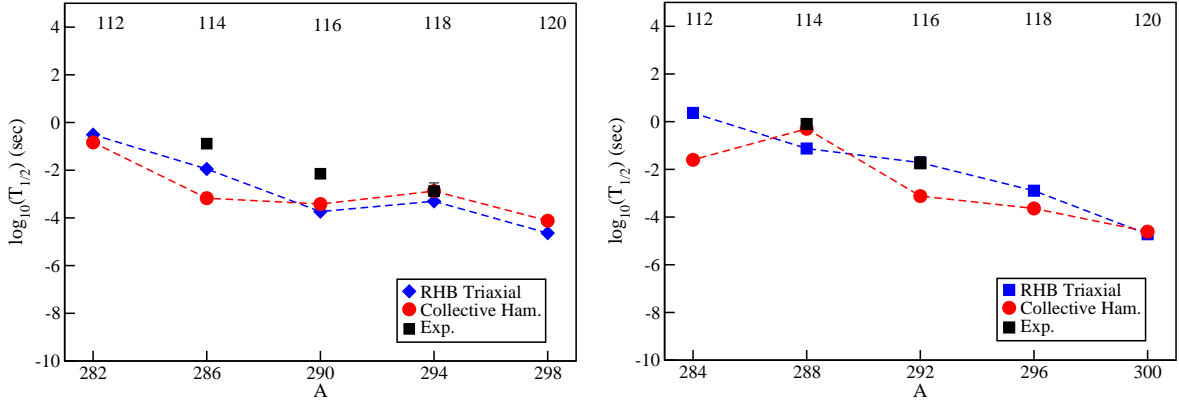

\includegraphics[scale=0.315]{Ta_116_290.eps}\quad
\includegraphics[scale=0.315]{Ta_116_292.eps}
\caption{(Color online)  Half-lives for the $\alpha$-decay chains of $^{298}$120 (left) and 
$^{300}$120 (right). The theoretical values are calculated from a phenomenological Viola-Seaborg-type formula 
\cite{ParSob.05,SobPom.07}, using the $Q_\alpha$ values from Fig.~\ref{qeven}. Diamonds correspond to 
values of $Q_\alpha$ calculated from mean-field RHB solutions, whereas circles denote half-lives computed  
using  $Q_\alpha$ values determined by the $0^+$ ground states of the quadrupole collective Hamiltonian. 
The data (squares) are from Ref.~\cite{og2007}.}
\label{teven}
\end{figure}
%
Alpha-decay half-lives are calculated using a simple five-parameter phenomenological Viola-Seaborg-type formula 
\cite{ParSob.05,SobPom.07}. The parameters of this formula were adjusted to experimental half-lives and 
$Q_\alpha$ values of more than 200 nuclei with $Z = 84-111$ and $N=128-161$ \cite{ParSob.05}. 
Using the theoretical $Q_\alpha$ values plotted in Fig.~\ref{qeven} as input, the resulting half-lives are compared 
to available data \cite{og2007} in Fig.~\ref{teven}. A rather good agreement with experiment 
is obtained for both decay chains.

The theoretical values denoted by (blue) diamonds in Figs.~\ref{qeven} and \ref{teven} correspond to transitions 
between the self-consistent mean-field minima on the triaxial RHB energy surfaces shown in Figs. \ref{pes1} 
and \ref{pes2}. Such a calculation does not explicitly take into account collective correlations related to symmetry
restoration and to fluctuations in the collective coordinates $\beta$ and $\gamma$. Physical transitions occur, of 
course, not between mean-field minima but between states with definite angular momentum. In cases in which 
both the initial and final states have similar deformation this will not make a large difference for the $Q_\alpha$ 
values, because collective correlations are implicitly taken into account in energy density functionals through the adjustment of parameters to ground-state properties (masses). The difference, however, can be larger in cases 
when the equilibrium shapes of the parent and daughter nucleus correspond to rather different deformations, 
because the collective correlation energy generally increases with deformation. For this reason we have also 
used a recent implementation of the collective Hamiltonian based on relativistic energy density 
functionals \cite{Nik.09}, to calculate $\alpha$-transition energies between ground states of even-even nuclei 
($0^+ \rightarrow 0^+$ transitions). Starting from self-consistent single-nucleon orbitals, the corresponding 
occupation probabilities and energies at each point on the energy surfaces shown in Figs.~\ref{pes1} 
and \ref{pes2}, the mass parameters and the moments of inertia of the collective Hamiltonian are calculated as 
functions of the deformations $\beta$ and $\gamma$. The diagonalization of the hamiltonian 
yields excitation energies and collective wave functions that can be used to calculate various 
observables.  The (red) circles in Figs.~\ref{qeven} and \ref{teven} denote the $Q_\alpha$ values and half-lives, 
respectively, computed for transitions $0^+_{\rm g.s.} \rightarrow 0^+_{\rm g.s.}$ between eigenstates of the 
collective Hamiltonian. The differences with respect to mean-field values are not large, especially for the heaviest, 
weakly oblate deformed or spherical systems. For the lighter prolate and more deformed nuclei, the differences 
can be as large as the deviations from experimental values.  

%
\begin{figure}
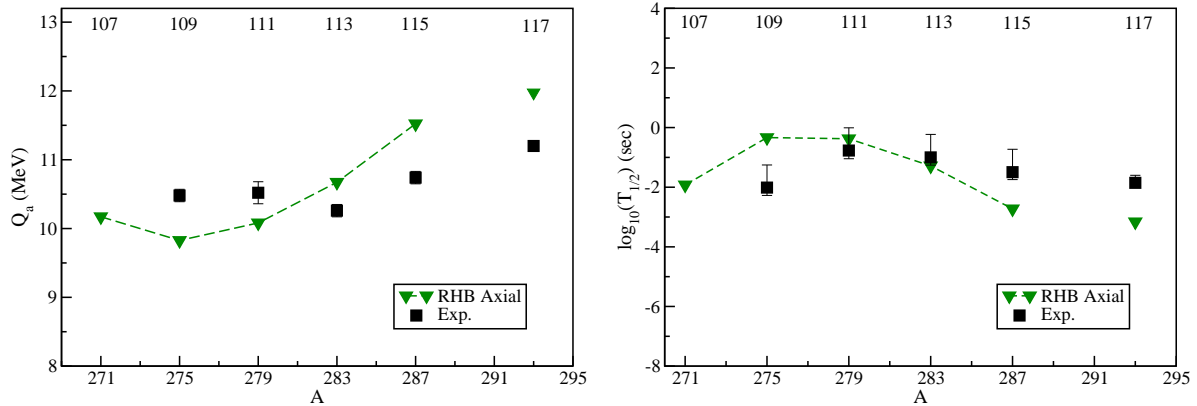

\includegraphics[scale=0.315]{Qa_odd_even.eps}\quad
\includegraphics[scale=0.315]{Ta_odd_even.eps}
\caption{(Color online) $Q_\alpha$ values (left), and half-lives (right) 
for the $\alpha$-decay chain of $^{287}$115, and for the nucleus 
$^{293}$117. }
\label{odd-even}
\end{figure}
%

%
\begin{figure}
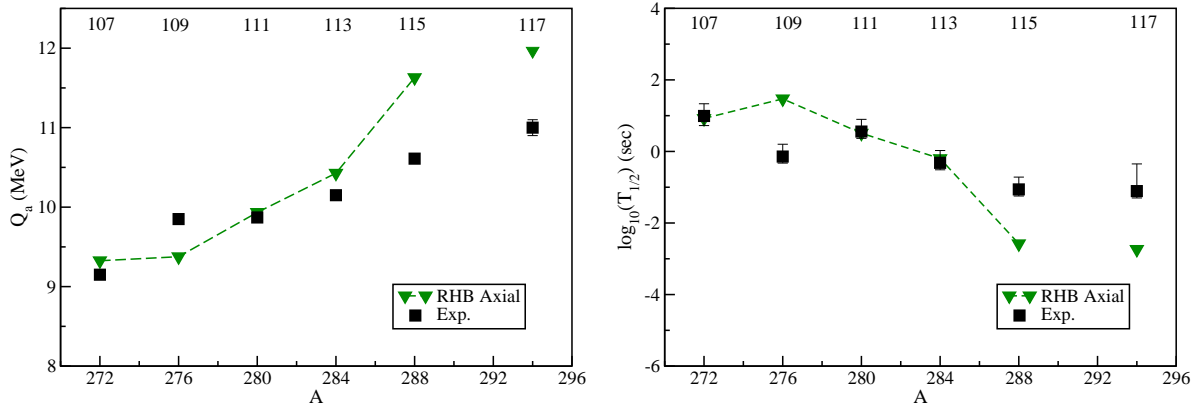

\includegraphics[scale=0.315]{Qa_odd_odd.eps}\quad
\includegraphics[scale=0.315]{Ta_odd_odd.eps}
\caption{(Color online) Same as in the caption to Fig.~\ref{odd-even} but for the 
$\alpha$-decay chain of $^{288}$115, and for the nucleus 
$^{294}$117.}
\label{odd-odd}
\end{figure}
%

Finally, in Figs.~\ref{odd-even} and \ref{odd-odd} we plot the $Q_\alpha$ values and half-lives for a chain of 
odd-even and odd-odd superheavy systems, respectively, in comparison with available data \cite{og2007,og2010}. 
The $Q_\alpha$ values are computed as energy differences between mean-field axial RHB minima of 
parent and daughter nuclei. The minima are determined in the blocking approximation for the odd proton (odd-even 
nuclei), and the odd proton and odd neutron (odd-odd nuclei). The equilibrium deformations are determined 
from the axial energy minima of the corresponding even-even systems, and the equilibrium odd-even (odd-odd) 
configuration is the one that minimizes the RHB total energy of the odd-even (odd-odd) system by blocking the  
odd-proton (odd-proton and odd-neutron) Nilsson orbitals. Because one needs to block several Nilsson orbitals 
in order to find the energy minimum, the calculation for odd-even and odd-odd superheavy nuclei is 
restricted to configurations with axial symmetry. Also in this case the model reproduces the mass dependence 
of $Q_\alpha$ values and half-lives, and only for the heaviest nuclei with $Z=115$ and $Z=117$ we find 
significant differences with respect to data. The half-lives are calculated with the Viola-Seaborg-type formula 
of Refs.~\cite{ParSob.05,SobPom.07}, which takes into account the effect of the odd nucleons by reducing 
the transition energy with respect to $Q_\alpha$ by the average excitation energy of the daughter nucleus. 
This correction is necessary when considering ground-state to ground-state $Q_\alpha$ values, because the 
half-life is determined by the most probable transition and this occurs between states with the 
same structure (same quantum numbers for odd-even and odd-odd nuclei). 
We note that a somewhat better agreement with data for $^{293}$117 
and $^{294}$117 was obtained using the HN macro-micro model that includes a more realistic 
deformation space \cite{Sob.10}. From the energy surfaces shown in Figs.~\ref{pes1} and \ref{pes2}, 
one notes that the corresponding even-even systems are soft both in $\beta$ and $\gamma$ and, therefore, 
pronounced effects of core polarization can be expected in the odd-even and odd-odd nuclei. The detailed 
structure of these soft nuclei cannot, of course, be reproduced by the simple blocking approximation 
assuming axially symmetric shapes, as used in the present calculation. Nevertheless, the level of agreement 
with data on the $Q_\alpha$ values and half-lives for odd-even and odd-odd nuclei reflects the underlying 
structure and ordering of proton and neutron quasiparticle states predicted by the functional DD-PC1 and the 
separable pairing interaction. 

\section{\label{secIV} Summary and outlook}

The framework of relativistic nuclear energy density functionals (EDFs) has been applied 
to a study of deformation effects and shapes of superheavy nuclei. The microscopic 
self-consistent calculation is based on the EDF DD-PC1 \cite{NVR.08}, and a separable 
pairing interaction, used in the relativistic Hartree-Bogoliubov (RHB) model. 

In addition to the equation of state of symmetric and asymmetric nuclear matter, the 
functional DD-PC1 was adjusted exclusively to the experimental masses of 64 axially 
deformed nuclei in the regions $A\approx 150-180$ and $A\approx 230-250$.  It is 
therefore interesting to note that, in self-consistent constrained calculations that include 
axially symmetric, triaxial, and reflection-asymmetric shapes, this functional 
reproduces not only the empirical masses, $Q_{\alpha}$ values, and equilibrium 
quadrupole deformations of actinide nuclei, but also the heights of the first and second 
fission barriers, as well as excitation energies of fission isomers. This is an important 
result as many modern functionals or effective interactions, used in studies of 
superheavy nuclei, are specifically adjusted to data on fission barriers and 
fission isomers. 

After testing the theoretical framework in the actinide region, we have performed 
a self-consistent RHB calculation of triaxial shapes for two $\alpha$-decay chains 
superheavy nuclei, starting from $^{298}$120 and $^{300}$120. In 
both chains the heaviest systems display soft oblate axial shapes with 
minima that extend from the spherical configuration to $|\beta_{20}| \approx 0.4$ 
($Z=120$) and  $|\beta_{20}| \approx 0.3$ ($Z=118$). The intermediate nuclei 
with $Z=116$ are essentially spherical but soft both in $\beta$ and $\gamma$, 
whereas prolate deformed mean-field minima develop in the lighter systems with $Z=114$, $Z=112$ 
and $Z=110$. The theoretical $Q_\alpha$ values reproduce the trend of the data, 
with the largest difference between theoretical and experimental values of less than 1 MeV.
Alpha-decay half-lives are calculated using a five-parameter phenomenological Viola-Seaborg-type 
formula, and a good agreement with experiment is obtained for both decay chains.
The $Q_\alpha$ values and half-lives for two chains of odd-even and odd-odd superheavy systems, 
have been computed using a simple blocking approximation in axially symmetric self-consistent 
RHB calculations. The theoretical values are in rather good agreement with the experimental 
$Q_\alpha$s and half-lives, and only for the heaviest nuclei with $Z=115$ and $Z=117$ one finds
significant differences, most probably caused by the restriction to axially symmetric shapes.

For the two chains of even-even superheavy nuclei we have also explicitly considered 
collective correlations related to symmetry restoration and fluctuations in the 
quadrupole collective coordinates. These correlations are implicitly taken into account 
when adjusting energy density functionals to binding energies, but their explicit 
treatment could be important in cases when the initial and final states of an $\alpha$-decay 
have markedly different deformations or shapes. Using a collective quadrupole Hamiltonian 
based on REDFs, we have calculated $\alpha$-transition energies between ground states of 
even-even nuclei ($0^+ \rightarrow 0^+$ transitions), rather than between mean-field 
minima. The resulting $Q_\alpha$ values and half-lives do not significantly differ from 
the corresponding mean-field values, except for lighter prolate and more deformed nuclei, 
for which the differences can be as large as the deviations from experimental values.

Together with the recent RMF+BCS study of fission barriers and fission paths of Ref.~\cite{Abu.12}, 
this work has demonstrated the potential of the new class of semi-empirical 
REDFs for studies of shape coexistence and triaxiality in the heaviest nuclear system, 
including the explicit treatment of collective correlations using a microscopic 
collective Hamiltonian. This opens the possibility for a more detailed analysis of this region 
of SHN, including all presently known nuclides with $Z=110-118$, as well as spectroscopic 
studies of nuclei with $Z>100$.  

\bigskip
\begin{acknowledgements}

The authors are grateful to A. V. Afanasjev, P. M\"oller, and P. Ring for valuable discussions. 
This work was supported by the MZOS - project 1191005-1010, the
Croatian Science Foundation, and the Greek State Scholarships Foundation (I.K.Y.).
\end{acknowledgements}
\clearpage




\begin{thebibliography}{99}
 
\bibitem{Brack.72} M. Brack, J. Damgaard,  A. S. Jensen, H. C. Pauli, V. M. Strutinski,
	C. Y. Wong, Rev. Mod. Phys. 44, 320 (1972).
 \bibitem{MN.94} P. M\"oller and J. R. Nix, J. Phys. G: Nucl. Part. Phys. 20, 1681 (1994). 
 
 \bibitem{MNMS.95} P. M\"oller, J. R. Nix, W. D. Myers, and W. J. Swiatecki, At. Data 
 	Nucl. Data Tables 59, 185 (1995). 
	
\bibitem{Sob.94} A. Sobiczewski, Phys. Part. Nucl. 25, 119 (1994). 

\bibitem{SobPom.07} A. Sobiczewski and K. Pomorski, Prog. Part. Nucl. Phys. 58, 292 (2007).

\bibitem{cwiok96}
S. \' Cwiok, J. Dobaczewski, P.-H. Heenen, Magierski, W. Nazarewicz,  Nucl. Phys. A 611, 211 (1996).

\bibitem{lala.96} G.A. Lalazissis, M.M. Sharma, P. Ring, and Y.K. Gambhir, 
	Nucl. Phys. A 608, 202 (1996).

\bibitem{Ben.98} M. Bender, K. Rutz, P.-G. Reinhard, J. A. Maruhn, and W. Greiner,
	Phys. Rev. C 58, 2126 (1998).
	
\bibitem{Ben.99} M. Bender, K. Rutz, P.-G. Reinhard, J. A. Maruhn, and W. Greiner,
	Phys. Rev. C 60, 034304 (1999).

\bibitem{cwiok99}
 S. \' Cwiok, W. Nazarewicz, and P. H. Heenen, Phys. Rev. Lett. 83, 1108 (1999). 
 
\bibitem{Ber.01}  J.-F. Berger, L. Bitaud, J. Decharg\' e, M. Girod, and K. Dietrich, 
	Nucl. Phys. A 685, 1 (2001).

\bibitem{War.02} M. Warda, J. L. Egido, L. M. Robledo, and K. Pomorski, 
	Phys. Rev. C 66, 014310 (2002).

\bibitem{HN.02} P.-H. Heenen and W. Nazarewicz, Europhys. News 33, 1 (2002).

\bibitem{Gor.02} S. Goriely, M. Samyn, P.-H. Heenen, J. M. Pearson, and F. Tondeur,
	 Phys. Rev. C 66, 024326 (2002).

\bibitem {BHR.03} M. Bender, P.-H. Heenen, and P.-G. Reinhard, Rev. Mod. Phys.
{75}, 121 (2003).

\bibitem{Bur.04} T. B\" urvenich, M. Bender, J. A. Maruhn, and P.-G. Reinhard, 
	Phys. Rev. C 69, 014307 (2004).

\bibitem{CHN.05} S. \' Cwiok, P.-H. Heenen,  W. Nazarewicz, Nature 433, 705 (2005). 

\bibitem{DGGL.06} J.-P. Delaroche, M. Girod, H. Goutte, J. Libert, Nucl. Phys. A 771, 103 (2006). 

\bibitem{Sheikh.09} J. A. Sheikh, W. Nazarewicz, and J. C. Pei, Phys. Rev. C 80, 011302 (2009).

\bibitem{Erler.10} J. Erler, P. KlŸpfel, and P.-G. Reinhard, Phys. Rev. C 82, 044307 (2010) 

\bibitem{LA.11} E. V. Litvinova and A. V. Afanasjev, Phys. Rev. C 84, 014305 (2011).

\bibitem{Lit.12} Elena Litvinova, Phys. Rev. C 85, 021303 (2012).

\bibitem{Abu.12} H. Abusara, A. V. Afanasjev, and P. Ring, Phys. Rev. C 85, 024314 (2012).

\bibitem{og2007} Yu. Ts. Oganessian, J. Phys. G: Nucl. Part. Phys. 34, R165 (2007).

\bibitem{hofmann2007} S. Hofmann et al., Eur. Phys. J. A 32, 251 (2007).

\bibitem{og2010} Yu. Ts. Oganessian et al., Phys. Rev. Lett. 104, 142502 (2010).

\bibitem{og2011} Yu. Ts. Oganessian et al., Phys. Rev. C 83, 054315 (2011).

\bibitem{og2012} Yu. Ts. Oganessian et al., Phys. Rev. Lett. 108, 022502 (2012).

\bibitem{stavsetra2009} L. Stavsetra et al., Phys. Rev. Lett. 103, 132502 (2009).

\bibitem{dullmann2010} Ch. E. D\"ullmann et al., Phys. Rev. Lett. 104, 252701 (2010).

\bibitem{ellison2010} P. A. Ellison, et al., Phys. Rev. Lett. 105, 182701 (2010).

\bibitem{JKS.11} P. Jachimowicz, M. Kowal, and J. Skalski, Phys. Rev. C 83, 054302 (2011).

\bibitem{NVR.08}  T. Nik\v si\' c, D. Vretenar, and P. Ring, Phys. Rev. C 78, 034318 (2008). 

\bibitem{Kor.10} M. Kortelainen, T. Lesinski, J. Mor\' e, W. Nazarewicz, J. Sarich, N. Schunck, 
	M. V. Stoitsov, and S. Wild, Phys. Rev. C 82, 024313 (2010). 

\bibitem{Kor.12} M. Kortelainen, J. McDonnell, W. Nazarewicz, P.-G. Reinhard, J. Sarich, 
	N. Schunck, M. V. Stoitsov, and S. M. Wild, Phys. Rev. C 85, 024304 (2012).
	
\bibitem{VALR.05} D. Vretenar, A.~V. Afanasjev, G.A. Lalazissis, and P. Ring, 
		Phys. Rep.  409, 101 (2005).

\bibitem {BGG.91}J.~F. Berger, M. Girod, and D. Gogny, Comp.
	Phys. Comm. 63, 365 (1991).

\bibitem {TMR.09a}Y. Tian, Z.~Y. Ma, and P. Ring, Phys. Lett. B 676, 44 (2009).

\bibitem{NRV.10} T. Nik\v{s}i\'{c}, P. Ring, D. Vretenar, Y. Tian, and Z.~Y. Ma,
				Phys. Rev. C 81, 054318 (2010).	
				
\bibitem{AW.03} G. Audi, A. H. Wapstra, and C. Thibault,
	Nucl. Phys. A 729, 337 (2003).	

\bibitem{RNT.01} S. Raman, C.W. Nestor, Jr., and P. Tikkanen, 
	At. Data Nucl. Data Tables 78, 1 (2001).

\bibitem{BBMR.04}
    T. B\"{u}rvenich, M. Bender, J. A. Maruhn, and P.-G. Reinhard,
     Phys. Rev.  C 69,  014307  (2004).

\bibitem{BQS.04} L. Bonneau, P. Quentin, and D. Samsoen,
    Eur. Phys. J. A 21, 391 (2004).

\bibitem{LGM.06} Hong-Feng L\" u, Li-Sheng Geng, and Jie Meng,
    Chin. Phys. Lett. 23, 2940 (2006).

\bibitem{Abu.10} H. Abusara, A. V. Afanasjev, and P. Ring, Phys. Rev. C 82, 044303 (2010).

\bibitem{Lu.12} Bing-Nan Lu, En-Guang Zhao, and Shan-Gui Zhou, Phys. Rev. C 85, 011301 (2012).

\bibitem{BMM.02}T. B{\"u}rvenich, D. G. Madland, J. A. Maruhn, and P.-G. Reinhard,
		Phys. Rev. C 65,  044308  (2002).

\bibitem{Li.10} Z. P. Li, T. Nik\v{s}i\'{c}, D. Vretenar, P. Ring, and J. Meng, Phys. Rev. C 81, 064321 (2010).

\bibitem{Capote} R. Capote, M. Herman, P. Oblo\v zinsk\' y, P. G. Young, S. Goriely, T. Belgya, A. V. Ignatyuk, 
A. J. Koning, S. Hilaire, V. A. Plujko, M. Avrigeanu, O. Bersillon, M. B. Chadwick, T. Fukahori, Zhigang Ge, 
Yinlu Han, S. Kailas, J. Kopecky, V. M. Maslov, G. Reffo, M. Sin, E. Sh. Soukhovitskii, and P. Talou, 
Nucl. Data Sheets 110, 3107 (2009); Reference Input Parameter Library (RIPL-3) [http://www-nds.iaea.org/RIPL-3/].

\bibitem{Moller2009}  P. M\"oller, A. J. Sierk, T. Ichikawa, A. Iwamoto, R. Bengtsson, H. Uhrenholt, and S. Aberg, 
					Phys. Rev. C 79, 064304 (2009).	

\bibitem{MPS.01} I Muntian, Z. Patyk, and A. Sobiczewski, Acta Phys. Pol. B 32, 691 (2001).

\bibitem{Sob.10} A. Sobiczewski, Acta Phys. Pol. B 41, 157 (2010).

\bibitem{Sob.11} A. Sobiczewski, Acta Phys. Pol. B 42, 1871 (2011).

\bibitem{ParSob.05} A. Parkhomenko and A. Sobiczewski, Acta Phys. Pol. B 36, 3095 (2005).

\bibitem{Nik.09}  T. Nik\v{s}i\'{c}, Z. P. Li, D. Vretenar,  L. Pr{\'o}chniak,
	J. Meng, and P. Ring, Phys. Rev. C 79, 034303 (2009).
	
\end{thebibliography}
\end{document}